\documentclass[12pt]{article}

\usepackage[T1]{fontenc}
\usepackage[utf8]{inputenc}
\usepackage{amsmath, amssymb, amsthm, mathtools}
\usepackage{graphicx}
\usepackage{booktabs}
\usepackage{placeins}
\usepackage{authblk}
\usepackage{array}
\usepackage{hyperref}
\usepackage{float}
\usepackage{xcolor}
\usepackage{listings}
\RequirePackage[round,authoryear]{natbib}
\usepackage{soul}
\newtheorem{remark}{Remark}
\usepackage{fancyhdr}
\usepackage{changes}
\usepackage{tabularx}
\usepackage{authblk}
\usepackage{fancyvrb}
\DefineVerbatimEnvironment{Code}{Verbatim}{
  fontsize=\small,
  frame=single
}

\newcommand{\pkg}[1]{\textsf{#1}}
\newcommand{\proglang}[1]{\textsf{#1}}
\newcommand{\code}[1]{\texttt{\detokenize{#1}}}

\title{FastJM: An \proglang{R} Package for Efficient Implementation of Semiparametric Joint Models for Longitudinal and Survival Data}

\author[1]{Shanpeng Li\textsuperscript{\(\dagger\)}}
\author[2]{Emily Ouyang\textsuperscript{\(\dagger\)}}
\author[2]{Ace Isabel Mejia-Sanchez}
\author[2]{Xinping Cui\textsuperscript{\(\ddagger\)}}
\author[3]{Gang Li\textsuperscript{\(\ddagger\)}}

\affil[1]{Department of Statistics and Data Science,
Beijing Normal--Hong Kong Baptist University, China}
\affil[2]{Department of Statistics,
University of California at Riverside, USA}
\affil[3]{Department of Biostatistics,
University of California at Los Angeles, USA}

\date{}
\begin{document}
\maketitle
\small
\noindent
\(\dagger\) These authors contributed equally and are co-first authors.\\
\(\ddagger\) These authors are co-corresponding authors.\\
Correspondence: Xinping Cui (xpcui@ucr.edu) and Gang Li (vli@ucla.edu).

\baselineskip=20.5pt

%% an abstract and keywords
\begin{abstract}
%\begin{abstract}
Joint models provide a flexible framework for characterizing the association between longitudinal and time-to-event processes and have been widely applied in biomedical research. However, fitting joint models can be computationally challenging for large-scale and complex biomedical data. This paper introduces the \proglang{R} package \pkg{FastJM}, which provides computationally efficient frequentist estimation for three classes of semiparametric joint models: joint models with a single longitudinal biomarker, joint models with multiple longitudinal biomarkers, and joint models with a single longitudinal biomarker with heterogeneous within-subject (WS) variability. Within an expectation--maximization framework, \pkg{FastJM} employs customized linear-scan algorithms to efficiently update the nonparametric baseline hazards, thereby addressing a major computational bottleneck in semiparametric joint modeling. The package also supports commonly used time-dependent latent association structures by integrating these algorithms with a landmark multivariate joint modeling framework. \pkg{FastJM} provides a unified interface for model specification, estimation, inference, visualization, dynamic prediction, and prediction performance assessment, including cross-validated time-dependent accuracy measures and time-independent concordance statistics. We describe the underlying methodology and software implementation and demonstrate the main functionality of \pkg{FastJM} through reproducible examples.

Keywords:  semiparametric joint modeling; longitudinal data; competing risks;
normal approximation; linear-scan algorithms; \proglang{R} package.
\end{abstract}
 %% without formatting
%% at least one keyword must be supplied

%% publication information
%% NOTE: Typically, this can be left commented and will be filled out by the technical editor
%% \Volume{50}
%% \Issue{9}
%% \Month{June}
%% \Year{2012}
%% \Submitdate{2012-06-04}
%% \Acceptdate{2012-06-04}

%% The address of (at least) one author should be given
%% in the following format:
%% It is also possible to add a telephone and fax number
%% before the e-mail in the following format:
%% Telephone: +43/512/507-7103
%% Fax: +43/512/507-2851

%% for those who use Sweave please include the following line (with % symbols):
%% need no \usepackage{Sweave.sty}

%% end of declarations %%%%%%%%%%%%%%%%%%%%%%%%%%%%%%%%%%%%%%%%%%%%%%%

%% include your article here, just as usual
%% Note that you should use the \pkg{}, \proglang{} and \code{} commands.

\section[Introduction]{Introduction}

Longitudinal biomarkers and time-to-event outcomes are commonly collected from the same participants and may be statistically dependent. They are often analyzed separately using a mixed-effects model for the longitudinal outcome and a survival model for the event-time outcome. Such separate analyses can be inadequate when a terminal event induces informative dropout in the longitudinal process \citep{elashoff2008joint,ElashoffLiLi17JMBook,henderson2000joint, Rizopoulos10JM,sattar2019joint} or when the survival model includes intermittently observed, error-prone time-varying covariates \citep{SongDavidianTsiatis02SemiparamJM,tsiatis2004joint,Wang06CorrectedScore,crowther2013joint}. Joint models address these settings by modeling the longitudinal and event processes simultaneously through shared latent structure \citep{henderson2000joint,tsiatis2004joint}. The simplest and most widely used formulation links one longitudinal biomarker to one time-to-event outcome  - for example, repeated systolic blood pressure measurements and the occurrence of a cardiovascular disease (CVD) event. The two submodels may be connected through shared random effects or through features of the latent longitudinal trajectory, such as its current value or slope. Modern studies, however, increasingly involve multiple  longitudinal biomarkers, recurrent  events,  competing  risks, and complex multilevel or high-dimensional data,  motivating more flexible joint-modeling methods.

Joint-modeling methods have expanded substantially over the past two decades, with software development progressing alongside these methodological advances. Early \proglang{R} packages, such as \pkg{JM} \citep{rizopoulos2010jm}, \pkg{JMbayes} \citep{rizopoulos2016r}, \pkg{joineR} \citep{hickey2018comparison}, and \pkg{JSM} \citep{xu2020semi}, primarily focused on joint models involving a single longitudinal biomarker and a time-to-event outcome. \pkg{JM} and \pkg{joineR} also accommodate competing risks outcomes, in which individuals may experience one of several mutually exclusive event types, such as disease progression, organ transplantation, or death, thus allowing investigators to examine how longitudinal biomarkers are associated with distinct cause-specific event risks rather than combining different event types into a single composite endpoint \citep{Rizopoulos10JM,ElashoffLiLi17JMBook}. More recent software supports multivariate longitudinal processes including 
%Another important development extends the framework to jointly model multiple correlated longitudinal biomarkers together with a time-to-event outcome, providing a more comprehensive characterization of the latent associations between the longitudinal and survival processes. Frequentist implementations include 
\pkg{joineRML} \citep{hickey2018joinerml}, \pkg{gmvjoint} \citep{murray2022fast,murray2023fast}, \pkg{JMbayes2} \citep{rizopoulos2024optimizing}, \pkg{rstanarm} \citep{rstanarm}, and \pkg{INLAjoint} \citep{rustand2024fast}. In addition, joint modeling has been extended to incorporates heterogeneous within-subject (WS) biomarker variability through mixed-effects location-scale models \citep{hedeker2008application}, since subject-specific variability may provide prognostic information beyond the mean longitudinal trajectory and improve both risk prediction and our understanding of disease progression; this extension is implemented in \pkg{FlexVarJM} \citep{courcoul2025location}. Collectively, these packages provide flexible tools for fitting an increasingly broad range of joint models arising in biomedical research, and their distinctive features are summarized in Table~\ref{tab:jm_packages}. 
\begin{table}[t!]
\centering
\caption{Comparison of the \proglang{R} packages for joint modeling.}
\label{tab:jm_packages}
\footnotesize
\setlength{\tabcolsep}{4pt}
\renewcommand{\arraystretch}{1.12}
\begin{tabularx}{\linewidth}{
    @{}
    l
    >{\centering\arraybackslash}p{2cm}
    >{\centering\arraybackslash}p{0.65cm}
    >{\raggedright\arraybackslash}p{4cm}
    >{\raggedright\arraybackslash}X
    @{}
}
\toprule
Package &
Longitudinal outcomes &
CR &
Estimation procedure&
Distinctive features$^{2}$ \\
\midrule
\pkg{JM}          & One      & Yes & Frequentist EM       & Flexible latent associations \\
\pkg{joineR}      & One      & Yes & Frequentist EM       & Bootstrap SE estimation \\
\pkg{JSM}      & One      & No & Frequentist EM       & Transformation survival models \\
\pkg{JMbayes}     & One      & No  & Bayesian MCMC     & Bayesian prediction \\
\pkg{FlexVarJM}   & One      & Yes & Frequentist ML   & Heterogeneous WS variability \\
\pkg{joineRML}    & Multiple & No  & MCEM     & Bootstrap/profile-likelihood SE estimation \\
\pkg{gmvjoint}    & Multiple & No  & Frequentist approximate EM & Generalized longitudinal outcomes \\
\pkg{JMbayes2}    & Multiple & Yes & Bayesian MCMC & Generalized longitudinal outcomes \\
\pkg{rstanarm}$^{1}$    & Multiple & No  & Stan-based Bayesian & MCMC /variational approximations/ optimization \\
\pkg{INLAjoint}   & Multiple & Yes & Approximate Bayesian by INLA     & Generalized longitudinal outcomes/flexible latent associations \\
\bottomrule
\end{tabularx}

\vspace{2pt}
\begin{minipage}{\linewidth}
\footnotesize
\textit{Notes:} CR denotes competing risks;
EM, expectation--maximization; SE, standard error; MCEM, Monte Carlo EM; MCMC, Markov
chain Monte Carlo; INLA, integrated nested Laplace approximation; ML, maximum likelihood; and WS, within-subject. $^{1}$\pkg{rstanarm} is built upon Stan \citep{carpenter2017stan}, a probabilistic programming language for full Bayesian modeling. $^{2}$The final column highlights selected some distinctive features related to
model specification, uncertainty quantification, or computation. It is
not intended to provide an exhaustive list of package functionality or
a ranking of the packages.\end{minipage}
\end{table}

Despite these methodological and computational advances, computational scalability remains a central obstacle to fitting complex joint models. Large electronic health records (EHRs) and biobank studies may include tens or hundreds of thousands of participants, repeated measurements, several biomarkers, and competing-risks events. As the number of biomarker-specific random effects increases, numerical integration becomes increasingly expensive, while repeated risk-set calculations added substantial cost as the sample size grows \citep{li2022efficient}. These challenges motivate \pkg{FastJM}  that scales efficiently in both dimensions. \pkg{FastJM} provides frequentist estimation for three classes of semiparametric joint models: \code{jmcs()} fits joint models with one longitudinal biomarker  and competing-risks event outcomes \citep{li2022efficient}; \code{mvjmcs()} fits multiple correlated biomarkers with competing-risks event outcomes \citep{li2025efficient}; and \code{JMMLSM()} incorporates heterogeneous WS variability through a mixed-effects location-scale submodel \citep{hedeker2008application}, with the WS variability parameters estimated within the EM framework \citep{li2023joint}. 
All three functions use EM-type estimation. Customized linear-scan algorithms developed by \citet{li2022efficient} are employed to update the nonparametric cause-specific baseline hazards without repeatedly evaluating every risk set. \code{jmcs()} and \code{JMMLSM()} evaluate the E-steps using Gauss–Hermite quadrature rule, whereas \code{mvjmcs()} uses a normal approximation proposed by \citep{murray2022fast} to avoid tensor-product quadrature in high dimensions \citep{li2025efficient}. These strategies are intended to improve scalability while retaining adequate numerical accuracy. Beyond model estimation, \pkg{FastJM} provides an integrated workflow for data visualization, model diagnostics, dynamic prediction, and assessment of predictive performance.

%All three modeling functions are implemented within a unified expectation-maximization (EM) framework. Computational efficiency is achieved by combining customized linear-scan algorithms for efficiently updating the nonparametric cause-specific baseline hazards with efficient approximations for evaluating the high-dimensional integrals arising in the E-step. Specifically, \code{jmcs()} and \code{JMMLSM()} employ the linear-scan algorithm developed by \citet{li2022efficient}, while \code{mvjmcs()} further incorporates the normal approximation proposed by \citet{murray2022fast} to efficiently approximate the high-dimensional integrals associated with multiple correlated random effects \citep{li2025efficient}. Together, these computational strategies substantially improve scalability while maintaining estimation accuracy relative to existing software.

The three core modeling functions in \pkg{FastJM} adopt the shared random-effects parameterization, which enables the scalable estimation algorithms implemented in the package. Consequently, they assume a time-invariant association structure defined through subject-specific random effects rather than the more computationally demanding time-dependent latent association structures commonly used in the joint modeling literature. To further broaden the modeling capabilities of \pkg{FastJM}, we introduce a landmark multivariate joint modeling approach built upon \code{mvjmcs()}. At a prespecified landmark time $s$, this approach fits a joint model among subjects who remain event-free and relates the post-landmark hazard to a latent biomarker summary evaluated at time $s$, such as the fitted current value or the random-effects contribution at that time. Because the biomarker summary remains fixed over the prediction window, this approach is computationally simpler than a conventional joint model in which the hazard depends on an evolving latent trajectory.

%Specifically, the proposed framework fits a multivariate joint model using subjects who remain event-free at a user-specified landmark time, while allowing the survival submodel to incorporate flexible latent association structures based on the latent longitudinal process evaluated at the landmark time. This strategy preserves the computational scalability of the proposed estimation algorithms while accommodating a broader class of latent association structures between the longitudinal and survival processes.

While joint modeling has also been extended to accommodate alternative outcome types, including binary, ordinal, or Poisson longitudinal responses \citep{murray2023fast,rizopoulos2024optimizing,rustand2024fast}, panel-count data \citep{zhou2017joint}, interval-censored data \citep{courcoul2026joint}, and recurrent events \citep{kim2012joint}, %These extensions involve distinct modeling and computational considerations. 
here we focus on joint models with one or more continuous longitudinal biomarkers modeled using Gaussian mixed-effects submodels and competing-risks event outcomes modeled using cause-specific proportional hazards submodels. The methodological developments, computational strategies, and software implementations presented here are tailored to this setting.

The remainder of this paper is organized as follows. Section~\ref{sec:2} reviews the statistical models and estimation methods underlying the three core modeling functions implemented in \pkg{FastJM}. Section~\ref{sec:3} introduces the package and its main functionalities. Section~\ref{sec:4} illustrates model fitting, inference, visualization, dynamic prediction, and prediction performance assessment. Section~\ref{sec:5} presents the landmark multivariate extension. Concluding remarks are given in Section~\ref{sec:6}. The appendices provide simulation settings and additional methodological and implementation details of the landmark multivariate joint modeling framework for the \pkg{FastJM} package.

\section{Models and Estimation Methods}
\label{sec:2}

\subsection{Notation and general joint modeling framework}
\label{Sec:2.1}

\pkg{FastJM} implements three classes of semiparametric joint models for longitudinal and time-to-event data: joint models with a single longitudinal biomarker, joint models with multiple longitudinal biomarkers, and joint models with a single longitudinal biomarker with heterogeneous within-subject (WS) variability. All three model classes follow the general joint modeling framework of \citet{wulfsohn1997joint}, in which the longitudinal and event processes are linked through subject-specific random effects. This section introduces the notation common to all three model classes before presenting their specific formulations and estimation procedures.

Suppose that each subject may experience one of $K$ distinct event types or be right censored during follow-up. Let $\widetilde{T}_i$ denote the event time, $\widetilde{D}_i \in \{1,\ldots,K\}$ the corresponding event type, and $C_i$ an independent, non-informative censoring time for subject $i$. The observed competing-risks data are
\[
(T_i,D_i)
=
\left\{
\min(\widetilde{T}_i,C_i),
\;
\widetilde{D}_i
I(\widetilde{T}_i\le C_i)
\right\},
\qquad i=1,\ldots,n.
\]

For the longitudinal process, let $Y_i(t)$ denote the longitudinal biomarker value for subject $i$ at time $t$. Measurements are observed at subject-specific, possibly irregular visit times $t_{ij}$, $j=1,\ldots,n_i$, where $n_i$ denotes the number of repeated measurements for subject $i$. The observed longitudinal data are therefore
\[
\{Y_i(t_{ij}),\; j=1,\ldots,n_i\},
\qquad i=1,\ldots,n.
\]

\subsection{Joint model classes}
\label{sec:JMclass}

\subsubsection{Joint models with a single longitudinal biomarker}
\label{sec:SJM}

We first consider the standard joint model with a single longitudinal biomarker, which serves as the baseline modeling framework implemented in \pkg{FastJM}. The longitudinal process is described by the linear mixed-effects submodel

\begin{eqnarray}
\label{eq1.1}
Y_i(t)
&=&
X_i^{\top}(t)\beta
+
Z_i^{\top}(t)b_i
+
\sigma\epsilon_i(t),\\
\epsilon_i(t)
&\stackrel{\mathrm{i.i.d.}}{\sim}&
\mathcal{N}(0,1),\cr
b_i
&\sim&
\mathcal{N}(0,\Sigma),\nonumber
\end{eqnarray}
where $\beta$ is the vector of fixed effects, $b_i$ is the subject-specific random effects vector with covariance matrix $\Sigma$, $X_i(t)$ and $Z_i(t)$ are the corresponding design vectors for the fixed and random effects, respectively. Conditional on $b_i$, the measurement errors $\epsilon_i(t)$ are assumed independent across visits, normally distributed with mean zero and constant variance $\sigma^2$, and independent of $b_i$.

The competing risks event process is represented by cause-specific Cox proportional-hazards submodels,

\begin{eqnarray}
\label{eq1.2}
\lambda_{ik}(t\mid W_i,b_i)
&=&
\lim_{h\rightarrow0}
\frac{
P(t\le\widetilde{T}_i<t+h,\,
\widetilde{D}_i=k
\mid
T_i\ge t,
W_i,
b_i)
}{h}
\cr
&=&
\lambda_{0k}(t)
\exp
\left(
W_i^{\top}\gamma_k
+
b_i^{\top}\alpha_k
\right),
\qquad
k=1,\ldots,K.
\end{eqnarray}

Here, $W_i$ denotes the vector of baseline covariates, $\gamma_k$ is the corresponding regression coefficient vector for cause $k$, $\lambda_{0k}(t)$ is an unspecified baseline cause-specific hazard function, and $\alpha_k$ is the vector of association parameters linking the longitudinal and event processes through the shared random effects. Thus, the two submodels are connected through $b_i$, with $\alpha_k$ quantifying the contribution of random effects on the cause-specific hazard for event type $k$.

This model forms the foundation of the \pkg{FastJM} framework. The next two model classes extend it by accommodating multiple longitudinal biomarkers and heterogeneous within-subject (WS) variability while retaining the same general joint modeling structure.

\subsubsection{Joint models with multiple longitudinal biomarkers}
\label{sec:MJM}

The joint model with multiple longitudinal biomarkers extends the framework introduced in Section~\ref{sec:SJM} by jointly modeling multiple correlated longitudinal biomarkers together with the competing-risks event process. Suppose subject $i$ has $G$ longitudinal biomarkers, where biomarker $g$ is measured at visit times $t_{ijg}$, $j=1,\ldots,n_{ig}$, for $g=1,\ldots,G$. The longitudinal submodels are

\begin{eqnarray}
\label{eq2.1}
Y_{ig}(t)
&=&
X_{ig}^{\top}(t)\beta_g
+
Z_{ig}^{\top}(t)b_{ig}
+
\sigma_g\epsilon_{ig}(t),\\
\epsilon_{ig}(t)
&\stackrel{\mathrm{i.i.d.}}{\sim}&
\mathcal{N}(0,1),\cr
b_i
&\sim&
\mathcal{N}(0,\Sigma). \nonumber
\end{eqnarray}

Here, $\beta_g$ and $b_{ig}$ denote the biomarker-specific fixed and random effects, respectively, and $X_{ig}(t)$ and $Z_{ig}(t)$ are the corresponding design vectors. The residual variance $\sigma_g^2$ may differ by biomarker but is constant within biomarkers across subjects and visits.
To account for the dependence among the biomarkers, the biomarker-specific random effects are combined into the joint random-effects vector
\[
b_i=(b_{i1}^{\top},\ldots,b_{iG}^{\top})^{\top},
\]
which is assumed to follow a multivariate normal distribution with covariance matrix
\[
\Sigma=
\left(
\begin{array}{ccc}
\Sigma_{11} & \cdots & \Sigma_{1G}\\
\vdots & \ddots & \vdots\\
\Sigma_{G1} & \cdots & \Sigma_{GG}
\end{array}
\right),
\]
where $\Sigma_{gg'}=\mathrm{cov}(b_{ig},b_{ig'})$ for $g,g'=1,\ldots,G$. The covariance matrix $\Sigma$ characterizes both the within-biomarker and between-biomarker dependence through the shared random-effects structure.

The competing-risks event process is modeled using the cause-specific Cox proportional hazards submodels,
\begin{eqnarray}
\label{eq2.2}
\lambda_{ik}(t\mid W_i,b_i)
&=&
\lambda_{0k}(t)
\exp\left(
W_i^{\top}\gamma_k
+
\sum_{g=1}^{G}
b_{ig}^{\top}\alpha_{gk}
\right),
\quad
k=1,\ldots,K.
\end{eqnarray}

Here, $\gamma_k$ denotes the regression coefficient vector for the baseline covariates, $\lambda_{0k}(t)$ is the unspecified baseline cause-specific hazard function, and $\alpha_{gk}$ is the vector of association parameters linking biomarker $g$ to the cause-specific hazard for event type $k$. The longitudinal and event processes are therefore linked through the joint random-effects vector $b_i$, allowing both within- and between-biomarker dependence to contribute to the association between the longitudinal biomarkers and the competing-risks outcome.

The joint model with a single longitudinal biomarker presented in Section~\ref{sec:SJM} is recovered as the special case when $G=1$.

\subsubsection{Joint models with a single longitudinal biomarker with heterogeneous WS variability}
\label{sec:WSVSJM}

The joint model with a single longitudinal biomarker with heterogeneous WS variability extends the framework introduced in Section~\ref{sec:SJM} by allowing the residual variability of the longitudinal outcome to vary across subjects and visit times. The longitudinal process is modeled using the following mixed-effects location-scale submodel:

\begin{eqnarray}
\label{eq3.1}
Y_i(t)
&=&
X_i^{\top}(t)\beta
+
Z_i^{\top}(t)b_i
+
\sigma_i(t)\epsilon_i(t),\\
\label{eq3.2}
\log\{\sigma_i^2(t)\}
&=&
U_i^{\top}(t)\tau
+
\omega_i,\\
\epsilon_i(t)
&\stackrel{\mathrm{i.i.d.}}{\sim}&
\mathcal{N}(0,1). \nonumber
\end{eqnarray}

Here, $\beta$ and $b_i$ denote the fixed and random effects for the longitudinal mean process, respectively, while $\tau$ and $\omega_i$ denote the corresponding fixed and random effects for the variance process. The design vectors $X_i(t)$, $Z_i(t)$, and $U_i(t)$ correspond to the fixed effects in the mean model, the random effects in the mean model, and the fixed effects in the variance model, respectively. The variance model allows the residual variance $\sigma_i^2(t)$ to vary across subjects and visit times, thereby accommodating heterogeneous WS variability.

\begin{remark}
A more general scale model could replace $\omega_i$ in submodel~\eqref{eq3.2} with $V_i^{\top}(t)\omega_i$, where $V_i(t)$ is a vector of time-dependent covariates associated with the random effects in the variance model. The current implementation of \code{JMMLSM()}, however, assumes a scalar random intercept $\omega_i$, corresponding to $V_i(t)\equiv1$.
\end{remark}

To account for the dependence between the longitudinal mean and variability processes, the random effects are combined into the joint vector
\[
\theta_i=(b_i^{\top},\omega_i)^{\top},
\]
which is assumed to follow a multivariate normal distribution,
\[
\theta_i\sim\mathcal{N}(0,\Sigma_{\theta}),
\]
with covariance matrix
\[
\Sigma_{\theta}
=
\begin{pmatrix}
\Sigma_{bb} & \Sigma_{b\omega}\\
\Sigma_{b\omega}^{\top} & \sigma_{\omega}^2
\end{pmatrix},
\]
where $\Sigma_{bb}=\mathrm{cov}(b_i,b_i)$,
$\sigma_{\omega}^2=\mathrm{var}(\omega_i)$, and
$\Sigma_{b\omega}=\mathrm{cov}(b_i,\omega_i)$.
The covariance matrix $\Sigma_{\theta}$ characterizes both the subject-specific longitudinal trajectories and the heterogeneous WS variability, together with their dependence through the shared random effects.

The competing-risks event process is modeled using the cause-specific Cox proportional hazards submodels,
\begin{eqnarray}
\label{eq3.3}
\lambda_{ik}(t\mid W_i,\theta_i)
&=&
\lambda_{0k}(t)
\exp\left(
W_i^{\top}\gamma_k
+
b_i^{\top}\alpha_k
+
\omega_i\nu_k
\right),
\quad
k=1,\ldots,K.
\end{eqnarray}

Here, $\gamma_k$ denotes the regression coefficient vector for the baseline covariates, $\lambda_{0k}(t)$ is the unspecified baseline cause-specific hazard function, $\alpha_k$ quantifies the association between the longitudinal mean trajectory and the event process through $b_i$, and $\nu_k$ quantifies the association between the subject-specific WS variability and the event process through $\omega_i$. Consequently, the longitudinal and event processes are linked through the joint random-effects vector $\theta_i$, allowing both the subject-specific mean trajectory and WS variability to be associated with the competing-risks outcome.

The joint model with a single longitudinal biomarker presented in Section~\ref{sec:SJM} is recovered as the special case when the scale submodel reduces to a constant residual variance and var($\omega_i$)=0.

\subsection{Maximum likelihood estimation and efficient implementations}
\label{sec:2.3}

\subsubsection{Maximum likelihood estimation}

Maximum likelihood estimation for joint models is based on the joint distribution of the longitudinal and time-to-event outcomes
$\{Y_i,T_i,D_i\}$ \citep{wulfsohn1997joint,tsiatis2004joint,
henderson2000joint}. We assume that, conditional on the covariates and subject-specific random effects, the longitudinal and event processes are independent. Under this assumption, the random effects capture both the latent association between the longitudinal and event outcomes and the within-subject correlation among repeated longitudinal measurements.

Let $\vartheta_i$ denote the complete vector of random effects for subject $i$, where $\vartheta_i=b_i$ for the joint models with a single or multiple longitudinal biomarkers and $\vartheta_i=\theta_i$ for the joint model with a single longitudinal biomarker with heterogeneous WS variability. Conditional on the covariates and $\vartheta_i$, the complete-data likelihood contribution for subject $i$ can be written as
\[
L_i(Y_i,T_i,D_i,\vartheta_i;\Psi)
=
f(Y_i\mid\vartheta_i;\Psi)
f(T_i,D_i\mid\vartheta_i;\Psi)
f(\vartheta_i;\Psi),
\]
where $\Psi$ denotes the full collection of unknown model parameters and baseline hazard functions, and the conditioning on observed covariates is omitted for notational simplicity.

For the joint models with a single longitudinal biomarker, with or without heterogeneous WS variability, the longitudinal component factorizes over repeated measurements as
\[
f(Y_i\mid\vartheta_i;\Psi)
=
\prod_{j=1}^{n_i}
f\{Y_i(t_{ij})\mid\vartheta_i;\Psi\}.
\]
For the joint model with multiple longitudinal biomarkers, it factorizes over both biomarkers and repeated measurements:
\begin{equation}
f(Y_i\mid\vartheta_i;\Psi)
=
\prod_{g=1}^{G}
\prod_{j=1}^{n_{ig}}
f\{Y_{ig}(t_{ijg})\mid\vartheta_i;\Psi\}.
\label{eq:longlik}
\end{equation}
Equation (9) gives the conditional competing-risks likelihood: the cause-specific hazard contributes when $D_i = k$, while censored observations contribute only through the overall survival term,
\begin{equation}
f(T_i,D_i\mid\vartheta_i;\Psi)
=
\left\{
\prod_{k=1}^{K}
\lambda_{ik}(T_i\mid\vartheta_i;\Psi)^{I(D_i=k)}
\right\}
\exp\left\{
-\sum_{k=1}^{K}
\int_0^{T_i}
\lambda_{ik}(u\mid\vartheta_i;\Psi)\,du
\right\}.
\label{eq:survlik}
\end{equation}
Integrating out the random effects gives the observed-data likelihood contribution
\[
L_i(Y_i,T_i,D_i;\Psi)
=
\int
f(Y_i\mid\vartheta_i;\Psi)
f(T_i,D_i\mid\vartheta_i;\Psi)
f(\vartheta_i;\Psi)\,d\vartheta_i,
\]
and the full observed-data likelihood is
\begin{equation}
L(\Psi)
=
\prod_{i=1}^{n}L_i(Y_i,T_i,D_i;\Psi).
\label{eq:obslik}
\end{equation}

Direct maximization of \eqref{eq:obslik} is computationally challenging because it requires integration over the random effects and estimation of the unspecified cause-specific baseline hazard functions. In addition, the cumulative hazard terms in \eqref{eq:survlik} generally require numerical evaluation. The three model-fitting functions in \pkg{FastJM} therefore use customized expectation--maximization (EM) algorithms, treating the subject-specific random effects as missing data.

At iteration $m$, the E-step evaluates
\begin{equation}
Q(\Psi\mid\Psi^{(m)})
=
E_{\Psi^{(m)}}\left[
\sum_{i=1}^{n}
\log L_i(Y_i,T_i,D_i,\vartheta_i;\Psi)
\,\middle|\,
Y_i,T_i,D_i
\right],
\label{eq:estep}
\end{equation}
where the expectation is taken with respect to the posterior distribution
\[
f(\vartheta_i\mid Y_i,T_i,D_i;\Psi^{(m)}).
\]
The M-step then updates the parameters according to
\begin{equation}
\Psi^{(m+1)}
=
\arg\max_{\Psi}
Q(\Psi\mid\Psi^{(m)}).
\label{eq:mstep}
\end{equation}
The E- and M-steps are repeated until the specified convergence criterion is satisfied.

%The E-step requires posterior expectations of functions of the random effects, whereas the M-step uses these expectations to update the model parameters and baseline hazard functions. Some M-step updates are available in closed form, while the remaining parameters are updated numerically. The principal computational challenges arise from evaluating posterior expectations when the random-effects dimension is large and from repeatedly computing event-time risk-set sums when updating the survival submodel. \pkg{FastJM} addresses these challenges through model-specific integration or approximation methods in the E-step and customized linear-scan algorithms in the M-step.

The estimation framework described above is common to all three joint model classes. Efficient implementation of the E-step and M-step is essential for large-scale applications because these steps require repeated numerical integration and repeated evaluations of event-time risk sets. The next subsection describes the computational strategies implemented in \pkg{FastJM} to address these challenges.

\subsubsection{Efficient implementation of the EM algorithm}

The efficient implementation of the EM algorithm in \pkg{FastJM} builds on the common estimation framework described above while adapting the computational strategy to the random-effects structure of each model. Below, we describe the efficient implementations developed by \citet{li2022efficient}, \citet{li2025efficient}, and \citet{li2023joint} for the joint models with a single longitudinal biomarker, multiple longitudinal biomarkers, and a single longitudinal biomarker with heterogeneous within-subject (WS) variability, respectively.

\paragraph{E-step.}
The principal task in the E-step is to evaluate conditional expectations of functions of $\vartheta_i$ with respect to
\[
f(\vartheta_i\mid Y_i,T_i,D_i;\Psi^{(m)}).
\]
The required expectations differ across the three model classes and generally do not have closed-form expressions. Accordingly, \pkg{FastJM} uses a computational strategy tailored to the random-effects structure of each model.

For the joint model with a single longitudinal biomarker, numerical integration can be performed using standard Gauss--Hermite quadrature. Standard quadrature may, however, require many quadrature points when the posterior distribution of the random effects is concentrated away from zero. \pkg{FastJM} therefore uses the pseudo-adaptive Gauss--Hermite quadrature approach described by \citet{rizopoulos2012fast}.
%\citet{rizopoulos2012joint}. 
The quadrature nodes are recentered and rescaled using subject-specific posterior modes and curvature matrices obtained from an initial linear mixed-effects model fit. These transformed nodes are computed before the EM iterations and subsequently held fixed, reducing the computational cost relative to fully adaptive quadrature while generally requiring fewer quadrature points than standard Gauss--Hermite quadrature.

For the joint model with multiple longitudinal biomarkers, the dimension of the joint random-effects vector can increase rapidly with the number of biomarkers. Tensor-product Gauss--Hermite quadrature becomes impractical in this setting because the number of quadrature nodes grows exponentially with the random-effects dimension. Monte Carlo EM methods may also be computationally demanding for large sample sizes. \pkg{FastJM} therefore adopts the normal approximation developed by 
\citet{bernhardt2015fast,murray2022fast,murray2023fast}.
%\citet{bernhardt2015joint} and \citet{murray2022fast,murray2023gmvjoint}. 
At each EM iteration, the posterior distribution of the random effects is approximated by a multivariate normal distribution centered at its posterior mode, with covariance determined by the local curvature of the log-posterior density. The required conditional moments are then evaluated using properties of the multivariate normal distribution, thereby avoiding high-dimensional numerical integration.

For the joint model with a single longitudinal biomarker and heterogeneous WS variability, the random-effects vector contains components governing both the longitudinal mean and the residual variance. Because the variance random effect enters the longitudinal likelihood through the logarithm of the residual variance, the corresponding posterior distribution can be more strongly non-Gaussian than that arising from the standard single-biomarker model. The pseudo-adaptive quadrature rule used by \code{jmcs()} is therefore not adopted for this model. Instead, \code{JMMLSM()} uses adaptive Gauss--Hermite quadrature \citep{naylor1982applications}, in which the quadrature nodes and weights are recentered and rescaled at each E-step according to the current posterior mode and curvature. This approach is more computationally intensive than pseudo-adaptive quadrature but provides greater flexibility for approximating the posterior distribution under the mixed-effects location-scale model.

Thus, \code{jmcs()}, \code{mvjmcs()}, and \code{JMMLSM()} use pseudo-adaptive Gauss--Hermite quadrature, a normal approximation, and adaptive Gauss--Hermite quadrature, respectively. Each method is selected to balance numerical accuracy and computational efficiency for the corresponding random-effects structure.

\paragraph{M-step.}
In the M-step, the model parameters and cause-specific baseline hazard functions are updated by maximizing the expected complete-data log-likelihood obtained from the E-step. Closed-form updates are available for several parameters in the longitudinal submodels and random-effects distributions. Parameters without closed-form solutions, including regression and association parameters in the survival submodels, are updated using a one-step Newton--Raphson procedure.

A major computational bottleneck arises when updating the survival submodel because the corresponding score functions and baseline hazard estimators involve sums over the subjects at risk at each distinct event time. A direct implementation recomputes these risk-set sums separately at every event time and therefore requires $O(n^2)$ operations in the worst case \citep{li2022efficient}.

When the covariates in the survival submodel are time independent, subjects can be ordered by observed event time and the required risk-set quantities can be computed recursively using cumulative sums. \pkg{FastJM} implements customized linear-scan algorithms that exploit this ordering and reuse previously computed quantities rather than reconstructing each risk set. These algorithms reduce the computational complexity of the relevant M-step calculations from $O(n^2)$ to $O(n)$ without changing the resulting parameter updates \citep{li2022efficient,li2025efficient}.

By combining model-specific approaches for evaluating the E-step with linear-scan algorithms for updating the survival component in the M-step, \pkg{FastJM} improves computational scalability with respect to both the sample size and the dimension of the random effects. These features are particularly important for fitting joint models to large biomedical datasets with repeated measurements, multiple biomarkers, and competing-risks outcomes.

\subsection{Standard error estimation and efficient computation}
In the semiparametric joint modeling framework, the baseline hazard functions are left completely unspecified. As a result, standard error estimation for the parametric components requires additional considerations. As discussed in \citet{elashoff2016joint} (Section 4.1 p.72), several approaches have been proposed in the literature, including the profile likelihood, observed information matrix, and bootstrap methods. In \pkg{FastJM}, we adopt the profile-likelihood approach due to its compatibility with the EM algorithm.

Let $\Omega$ denote the vector of the parametric component in $\Psi = \left\{\Omega, \lambda_{01}(\cdot), \ldots, \lambda_{0K}(\cdot)\right\}$, and let $\hat{\Omega}$ be its semiparametric maximum likelihood estimate. The variance–covariance matrix of $\hat{\Omega}$ is estimated by inverting the empirical Fisher information matrix based on the profile likelihood of $\Omega$ \citep{lin2004latent, zeng2005asymptotic, zeng2005simultaneous}:
\begin{equation}
\label{SEestimation}
\sum_{i=1}^n \left[\nabla_{\Omega} l^{(i)}(\hat{\Omega}; Y, T, D)\right]
\left[\nabla_{\Omega} l^{(i)}(\hat{\Omega}; Y, T, D)\right]^{\top},
\end{equation}
where $\nabla_{\Omega} l^{(i)}(\hat{\Omega}; Y, T, D)$ denotes the observed score vector for the $i$th subject obtained from the profile likelihood $l^{(i)}(\Omega; Y, T, D)$ after profiling out the baseline hazards.
 
It is worth noting that direct computation of \eqref{SEestimation} is computationally intensive because it involves nested summations over subjects and event-time risk sets, resulting in an $O(n^3)$ computational complexity \citep{li2022efficient}. To address this challenge, \pkg{FastJM} extends the linear scan algorithms developed for parameter estimation to efficiently evaluate the survival component of the profile-score function. By reusing cumulative quantities and eliminating redundant summations over risk sets, the computational complexity is reduced from $O(n^3)$ to $O(n)$. This efficient implementation makes profile-likelihood--based statistical inference computationally feasible for large-scale joint modeling applications.

\subsection{Dynamic prediction of competing risk time-to-event data}
\label{sec:DP}
Joint models not only offers a general framework to study the association between the longitudinal and competing-risk time-to-event outcomes, but also facilitates subject-level dynamic prediction of cumulative incidence probabilities. In particular, based on a fitted joint model, we are interested in predicting cumulative incidence probabilities for a new subject $i^*$ that has provided the past longitudinal history 
$Y_{i^*}^{(s)}=\{Y_{i^*}(t), t\leq s\}$ prior to a landmark time $s>0$ and that an event has yet to happen by the landmark time $s$,  the cumulative incidence probability for type $k$ failure at a horizon time $u>s$ is 
\begin{eqnarray}
\label{competingdy}
P_{i^* k}(u, s|\Psi) &=& \text{Pr}(T_{i^*} \leq u, D_{i^*} = k |T_{i^*} > s, Y_{i^*}^{(s)}; \Psi)\cr
%&=& \int \frac{\text{Pr}(T_{i^*} \leq h, D_{i^*} = k, T_{i^*} > s| \vartheta_{i^*}; \Psi)}{\text{Pr}(T_{i^*} > s| \vartheta_{i^*}; \Psi)}f(\vartheta_{i^*}|T_{i^*} > s, Y_{i^*}^{(s)}; \Psi) d\vartheta_{i^*} \cr
%&=& \frac{\int \frac{CIF_{i^*k}(u,s|\vartheta_{i^*}; \Psi)}{S_{i^*}(s|\vartheta_{i^*}; \Psi)} f(Y_{i^*}^{(s)}|\vartheta_{i^*}; \Psi)S_{i^*}(s|\vartheta_{i^*}; \Psi)f(\vartheta_{i^*} | \Psi) d\vartheta_{i^*}}{\int f(Y_{i^*}^{(s)}|\vartheta_{i^*}; \Psi)S_{i^*}(s|\vartheta_{i^*}; \Psi)f(\vartheta_{i^*} | \Psi) d\vartheta_{i^*}},\cr
&=& \frac{\int CIF_{i^*k}(u,s|\vartheta_{i^*}; \Psi) f(Y_{i^*}^{(s)}|\vartheta_{i^*}; \Psi)f(\vartheta_{i^*} | \Psi) d\vartheta_{i^*}}{\int f(Y_{i^*}^{(s)}|\vartheta_{i^*}; \Psi)S_{i^*}(s|\vartheta_{i^*}; \Psi)f(\vartheta_{i^*} | \Psi) d\vartheta_{i^*}},
\label{pikus}
\end{eqnarray}
where 
\[S_{i^*}(t|\vartheta_{i^*}; \Psi) = \exp\left\{
-\sum_{k=1}^K \int_0^{t} \lambda_{i^*k}(l \mid \vartheta_{i^*}; \Psi)\,dl
\right\},\]
is the overall survival function, and
\[CIF_{i^* k}(u,s|\vartheta_{i^*}; \Psi) = \int_s^u S_{i^*}(t|\vartheta_{i^*}; \Psi)\lambda_{i^*k}(t \mid \vartheta_{i^*}; \Psi)\,dt\]
is the cumulative incidence function (CIF) for type $k$ failure. An estimate of $P_{i^* k}(u, s|\Psi)$ is then obtained by replacing $\Psi$, $S_{i^*}(.)$, and $CIF_{i^* k}(.)$ with their sample estimates $\hat{\Psi}$, $\hat{S}_{i^*}(.)$, and $\widehat{CIF}_{i^* k}(.)$, respectively.

The prediction performance of a joint model for competing risks outcomes can be evaluated using cross-validated measures of prediction accuracy. Prediction accuracy can be assessed using metrics such as the mean absolute prediction error across quantiles of risk scores (MAEQ) \citep{li2023joint}, the Brier score \citep{wu2018quantifying}, and generalized $R^2$ \citep{li2016prediction, zhuang2025time}, while discrimination is commonly evaluated using the area under the receiver operating characteristic curve (AUC) \citep{blanche2013time} and the concordance index (C-index) \citep{wolbers2014concordance}.

\section{The \proglang{R} package FastJM}
\label{sec:3}
\subsection{Package Overview}

The \proglang{R} package \pkg{FastJM} provides computationally efficient tools for fitting joint models of longitudinal biomarkers and competing-risks time-to-event outcomes. The package currently implements three main model-fitting functions. The function \code{jmcs()} fits a joint model for a single longitudinal biomarker. The function \code{mvjmcs()} extends the framework of \code{jmcs()}, allowing multiple biomarkers to be jointly modeled together with a competing-risks event process. The function \code{JMMLSM()} is another extension of \code{jmcs()} by allowing WS variability to vary across visits and subjects and enabling assessment of its association with the risk of clinical events.

Although \code{jmcs()}, \code{mvjmcs()}, and \code{JMMLSM()} target different joint modeling settings, they are designed around a common formula-based interface, as shown in Table~\ref{tab:key-arguments}. Each function takes a longitudinal data frame \code{ydata} in long format and a survival data frame \code{cdata} with one observation per subject. Users specify the longitudinal and survival components through \code{long.formula} and \code{surv.formula}, respectively, and define the subject-specific random-effects structure through the \code{random} argument. \code{mvjmcs()} allows \code{long.formula} and \code{random} to be supplied as lists to accommodate multiple longitudinal biomarkers. \code{JMMLSM()} further extends this interface by including \code{variance.formula} to model heterogeneous WS variability of the longitudinal outcome.
\begin{table}[t]
\centering
\begin{tabular}{ll}
\hline
Argument & Role \\
\hline
\code{ydata} 
& Longitudinal data in long format \\

\code{cdata} 
& Subject-level survival data \\

\code{long.formula} 
& Longitudinal submodel specification \\

\code{surv.formula} 
& Competing-risks survival submodel specification \\

\code{random} 
& One-sided random-effects specification \\

\code{variance.formula} 
& One-sided within-subject variability submodel in \code{JMMLSM()} \\

\code{control} 
& List of numerical and algorithmic options for model fitting \\
\hline
\end{tabular}
\caption{Key input arguments in the three main model-fitting functions of \pkg{FastJM}.}
\label{tab:key-arguments}
\end{table}

\begin{figure}[!htbp]
    \centering
    \includegraphics[width=0.7\linewidth]{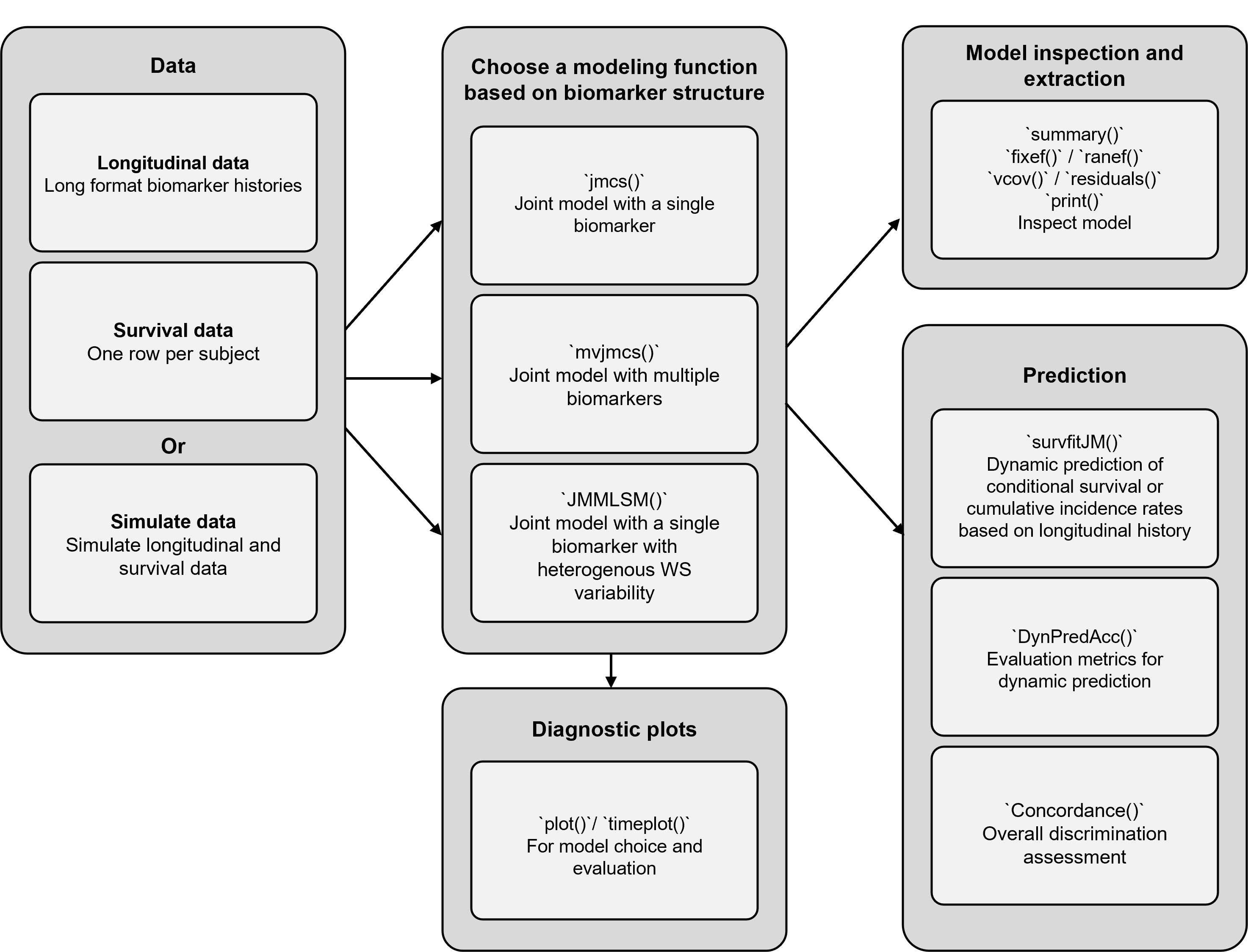}
    \caption{Overview of the \pkg{FastJM} workflow. Users supply longitudinal biomarker histories and subject-level survival outcomes, or generate both using the package’s data simulation tools. The appropriate model-fitting function is selected according to the number and structure of the biomarkers. Fitted models can then be inspected, visualized, and used for dynamic prediction and predictive-performance assessment.}
    \label{fig:flowchart}
\end{figure}

Figure~\ref{fig:flowchart} summarizes the overall workflow of \pkg{FastJM}, from data preparation and model fitting to post-estimation analysis. Users should provide longitudinal biomarker data in long format and subject-level survival data, or generate both types of data using the package's simulation tools. In the single-biomarker setting, \code{timeplot()} visualizes the longitudinal data to reveal potential heterogeneous WS variability, whereas \code{plot()} provides diagnostic plots for assessing the assumptions of a model fitted by \code{jmcs()}. The fitted models can also be inspected using standard summary and extraction methods and used for dynamic prediction and predictive-performance assessment. Further details about these post-estimation functions are provided in Section~\ref{sec:3.2}.

\subsection{Implementation details}
\label{sec:3.2}
The \code{control} argument of \code{jmcs()}, \code{mvjmcs()}, and \code{JMMLSM()} collects numerical and algorithmic options used in model fitting. Control settings are constructed using \code{jmcs_control()}, \code{mvjmcs_control()}, and \code{JMMLSM_control()}, respectively. Common options across the three control functions include \code{opt}, which specifies the optimizer used by the \pkg{nlme} package \citep{pinheiro2025nlme} when fitting the linear mixed-effects submodel to obtain initial parameter values, and the available options are \code{"nlminb"} and \code{"optim"};
\code{maxiter}, which specifies the maximum number of EM iterations; \code{tol}, which sets the convergence tolerance; \code{initial.para}, which accepts a list of initial parameter values for model fitting and defaults to \code{NULL}, in which case the longitudinal and survival submodels are fitted separately to obtain initial estimates; and \code{verbose}, which determines whether iterated parameter values are printed during model fitting. Further details on \code{maxiter} and \code{tol} are provided later in this section.

For \code{jmcs()} and \code{JMMLSM()}, the \code{control} argument also includes
\code{method} and \code{quadpoint}, which jointly determine how the integrals in the E-step are approximated. Specifically, \code{method} specifies the type of Gauss--Hermite quadrature rule, whereas \code{quadpoint} specifies the number
of quadrature points used under the selected rule. The available quadrature
options are:
\begin{list}{}{\leftmargin=1.5em \itemindent=0pt \itemsep=0pt
\topsep=0pt \partopsep=0pt \parsep=0pt \parskip=0pt}
    \item \code{method = "standard"}: standard Gauss--Hermite quadrature is used for both \code{jmcs()} and \code{JMMLSM()}.
    \item \code{method = "pseudo-adaptive"}: pseudo-adaptive Gauss--Hermite quadrature is used for \code{jmcs()}.
    \item \code{method = "adaptive"}: adaptive Gauss--Hermite quadrature is used for \code{JMMLSM()}.
\end{list}
The choice of \code{quadpoint} depends on the selected quadrature method and the dimension of the random effects. For standard Gauss--Hermite quadrature, a relatively large number of quadrature points is usually required to obtain accurate approximations. For example, at least 20 quadrature points are often needed for models with two-dimensional random effects. Because the total number of quadrature points increases rapidly with the dimension of the random effects, standard Gauss--Hermite quadrature is generally not recommended for models with high-dimensional random effects. Alternatively, adaptive and pseudo-adaptive Gauss--Hermite quadrature can often achieve satisfactory estimation accuracy with substantially fewer quadrature points because the integrand will be centered and scaled based on the posterior distribution of the random effects conditional on subject data. The default value for these two methods is set \code{quadpoint = 6}, which provides a practical balance between numerical accuracy and computational efficiency.

For \code{JMMLSM()}, the subject-specific quadrature calculations in the E-step can be performed independently. The implementation therefore supports parallel computation through the \code{cpu.cores} component of \code{control}, with \code{cpu.cores = 1} as the default. Users can specify the number of CPU cores used during estimation, which can further reduce the computation time for large datasets or higher-dimensional random effects.

For \code{mvjmcs()}, the E-step is computed using a normal approximation utilizing the subject-specific posterior distribution of the random effects. Consequently, \code{method} and \code{quadpoint} are not applicable in \code{control}, because quadrature-based numerical integration is not considered. This approximation substantially reduces the computational burden associated with high-dimensional random effects and facilitates efficient estimation in multivariate joint models. Likewise in \code{JMMLSM()}, \code{mvjmcs()} supports parallel computation through the \code{cpu.cores} component of \code{control}, with \code{cpu.cores = 1} as the default. Users may increase \code{cpu.cores} to reduce computation time for large datasets or models involving multiple longitudinal outcomes.

For all three model-fitting functions, estimation is based on an EM-type algorithm that is iterated until either the maximum number of iterations, specified by \code{maxiter}, is reached or convergence is declared. Naturally, a convergence criterion is needed to implement the algorithm
and in \pkg{FastJM}, convergence is declared whenever the relative difference between the parameter estimates from two consecutive iterations, that is, 
\begin{equation*}
\max \left\{
\frac{\left|\Psi^{(m)} - \Psi^{(m-1)}\right|}
{\left|\Psi^{(m-1)}\right| + 10 \times\,\texttt{tol}}
\right\}
< \texttt{tol}
\end{equation*}
is satisfied. The default value for \code{tol} is $10^{-4}$, which are the standard choices to claim convergence of iterative
algorithms. The component \code{maxiter} states the maximum number of iterations for the EM algorithm with the default value of 10000.

Several supporting functions are provided to extract or compute useful
quantities from fitted joint models, including model summaries, statistical inference for estimated coefficients, empirical Bayes estimates, and covariance matrices of parameter estimates. Commonly used S3 methods are available for objects returned by \code{jmcs()}, \code{mvjmcs()}, and \code{JMMLSM()}, including \code{print()}, \code{summary()}, \code{fixef()}, \code{ranef()}, and \code{vcov()}.

For objects of class \code{jmcs}, the \code{residuals()} method computes
model residuals based on the observed data, and the \code{plot()} method
provides diagnostic plots for the fitted joint model. Dynamic prediction of cumulative incidence probabilities, as described in Section~\ref{sec:DP}, is implemented through the generic function \code{survfitJM()}, which dispatches to the appropriate model-specific prediction routine for objects returned by \code{jmcs()}, \code{mvjmcs()}, or \code{JMMLSM()}. The \code{DynPredAcc()} function computes cross-validated measures of calibration and discrimination for dynamic predictions at specified landmark and horizon times, whereas
\code{Concordance()} calculates cross-validated concordance statistics
based on time-independent prognostic indices \citep{royston2013external}. Both functions support all
three classes of joint models implemented in \pkg{FastJM}. Detailed
descriptions of these functions are provided in the online help files.

\section{Illustrating examples}
\label{sec:4}
The three classes of joint models implemented in \pkg{FastJM} focus on continuous longitudinal outcomes modeled using Gaussian mixed-effects submodels and competing-risks outcomes modeled using cause-specific proportional hazards submodels. In this section, we illustrate each model class through a separate example with simulated datasets. Specifically, we demonstrate the joint models with a single longitudinal biomarker implemented in \code{jmcs()}, the joint models with multiple longitudinal biomarkers implemented in \code{mvjmcs()}, and the joint mixed-effects location-scale model implemented in \code{JMMLSM()}, together with model specification, estimation, and interpretation, as well as the supporting visualization, diagnostic, and prediction tools provided by \pkg{FastJM}.
\subsection{Fitting a joint model with a single longitudinal biomarker}
\label{sec:expjmcs}
We first simulate two datasets comprising repeated longitudinal biomarker measurements and competing-risks event data. For this purpose, \pkg{FastJM} provides \code{simJMdata()}, a data-generation function developed as part of the package for simulating data under the univariate joint modeling framework. Details of the arguments in \code{simJMdata()} and the corresponding generative joint model formulas are provided in Appendix \ref{appen:jmcs}. We used \code{simJMdata()} to generate the joint model data, and the resulting longitudinal dataset, \code{ydata}, contains repeated biomarker measurements over follow-up, whereas the survival dataset, \code{cdata}, contains the corresponding event times, competing event indicators, and baseline covariates.
\begin{Code}
R> library("FastJM")
R> dat <- simJMdata(...)
R> ydata <- dat[["ydata"]]
R> cdata <- dat[["cdata"]]
\end{Code}
\paragraph{Model fitting}
We fit the following joint model for the longitudinal trajectory and the event process, where the longitudinal submodel is modeled through a linear mixed effects model with homogeneous WS variability, and the event process is modeled through a cause-specific proportional hazards submodel, linked by the shared random effects:
\[
Y_i(t_{ij})
=
\beta_0
+ \beta_1 X_{i1}
+ \beta_2 X_{i2}
+ \beta_3 X_{i3}
+ \beta_4 t_{ij}
+ b_{i0}
+ b_{i1} t_{ij}
+ \epsilon_i(t_{ij}), \quad
\epsilon_i(t_{ij}) \sim N(0, \sigma^2),
\]
\[
\lambda_{ik}(t \mid  X_i, b_i)
=
\lambda_{0k}(t)
\exp\left( X_i^\top \gamma_k
+ b_i^\top \alpha_k
\right), \quad k = 1,2,
\]
\begin{Code}
R> fit <- jmcs(
+     ydata = ydata,
+     cdata = cdata,
+     long.formula = Y ~ X1 + X2 + X3 + time,
+     random =  ~ time | ID,
+     surv.formula = Surv(survtime, cmprsk) ~ X1 + X2 + X3,
+     control = jmcs_control(quadpoint = 3)
+  )
\end{Code}
The fitted model object contains parameter estimates for both the
longitudinal and survival submodels, together with the matrix of estimated
standard deviation and correlation among the random effects. The \code{print()}
method reports parameter estimates, standard errors, Wald test statistics, and
associated \emph{p}-values for the longitudinal and survival components.
The survival fixed effects and association parameters are labeled using
descriptive suffixes for ease of interpretation. In particular,
\code{\_1} and \code{\_2} denote failure types 1 and 2, respectively.
\begin{Code}
R> fit

Call:
 jmcs(ydata = ydata, cdata = cdata, long.formula = Y ~ X1 + X2 + X3 + time, 
 random = ~time | ID, 
 surv.formula = Surv(survtime, cmprsk) ~ X1 + X2 + X3, 
 control = jmcs_control(quadpoint = 3)) 

Data Summary:
Number of observations: 7250 
Number of groups: 1000 

Proportion of competing risks: 
Risk 1 : 48.9 %
Risk 2 : 36.6 %

Numerical intergration:
Method: pseudo-adaptive Guass-Hermite quadrature
Number of quadrature points:  3 

Model Type: joint modeling of longitudinal continuous and competing risks data 

Model summary:
Longitudinal process: linear mixed effects model
Event process: cause-specific Cox proportional hazard model with non-parametric 
baseline hazard

Loglikelihood:  -17773.33 

Fixed effects in the longitudinal sub-model:  Y ~ X1 + X2 + X3 + time 

            Estimate     SE Z value  p-val
(Intercept)   4.9469 0.0593 83.3887 0.0000
X1            1.5457 0.0775 19.9488 0.0000
X2            2.0028 0.0671 29.8323 0.0000
X3            1.0543 0.0197 53.5084 0.0000
time          1.9405 0.0418 46.4238 0.0000

Residual error:
         Variance StdDev
Residual   0.9869 0.9934

Fixed effects in the survival sub-model:  Surv(survtime, cmprsk) ~ X1 + X2 + X3 

     Estimate     SE Z value  p-val
X1_1   1.0106 0.1243  8.1292 0.0000
X2_1   0.4712 0.1010  4.6669 0.0000
X3_1   0.4920 0.0334 14.7371 0.0000
X1_2  -0.5735 0.1313 -4.3664 0.0000
X2_2   0.4210 0.1106  3.8050 0.0001
X3_2   0.1711 0.0344  4.9808 0.0000

Association parameters:                 
              Estimate     SE  Z value  p-val
(Intercept)_1   0.9593 0.0811  11.8329 0.0000
time_1          0.6172 0.0756   8.1628 0.0000
(Intercept)_2  -0.9169 0.0907 -10.1139 0.0000
time_2         -0.4772 0.0816  -5.8491 0.0000

Random effects:                 
  Formula: ~time | ID 
            StdDev  (Intr)
(Intercept) 1.0359        
time        1.0003 -0.0427
\end{Code}
\paragraph{Model diagnostics}
\pkg{FastJM} provides a standard diagnostic plotting method through \code{plot()}, which can be used to assess model fit. As shown below, \code{plot(fit)}, produced in Figure~\ref{fig:plotjmcs}, which displays residuals versus fitted values and a normal Q--Q plot for the longitudinal submodel, together with the estimated marginal survival and cumulative hazard functions for the event process. The marginal survival function is approximated by averaging the conditional survival functions over the empirical Bayes estimates of the random effects,
\(\widehat S(t) = \int \exp\left\{-\sum_{k=1}^{K}\int_0^{t}\lambda_{ik}(u \mid \vartheta_i; \hat{\Psi})du\right\} f(\vartheta_i; \hat{\Psi}) \, d\vartheta_i \approx n^{-1}\sum_{i=1}^n
\exp\{-\sum_{k=1}^K \int_0^{t}
\lambda_{ik}(u \mid \widehat{\vartheta}_i; \widehat{\Psi})\,du\}\),
where \(\hat{\vartheta}_i\) denotes the empirical Bayes estimates of the random effects. The marginal cumulative hazard function is calculated as \(\widehat \Lambda(t) = -\log \widehat S(t)\).
\begin{Code}
R> plot(fit)
\end{Code}
\begin{figure}[!ht]
    \centering
\includegraphics[width=0.7\linewidth]{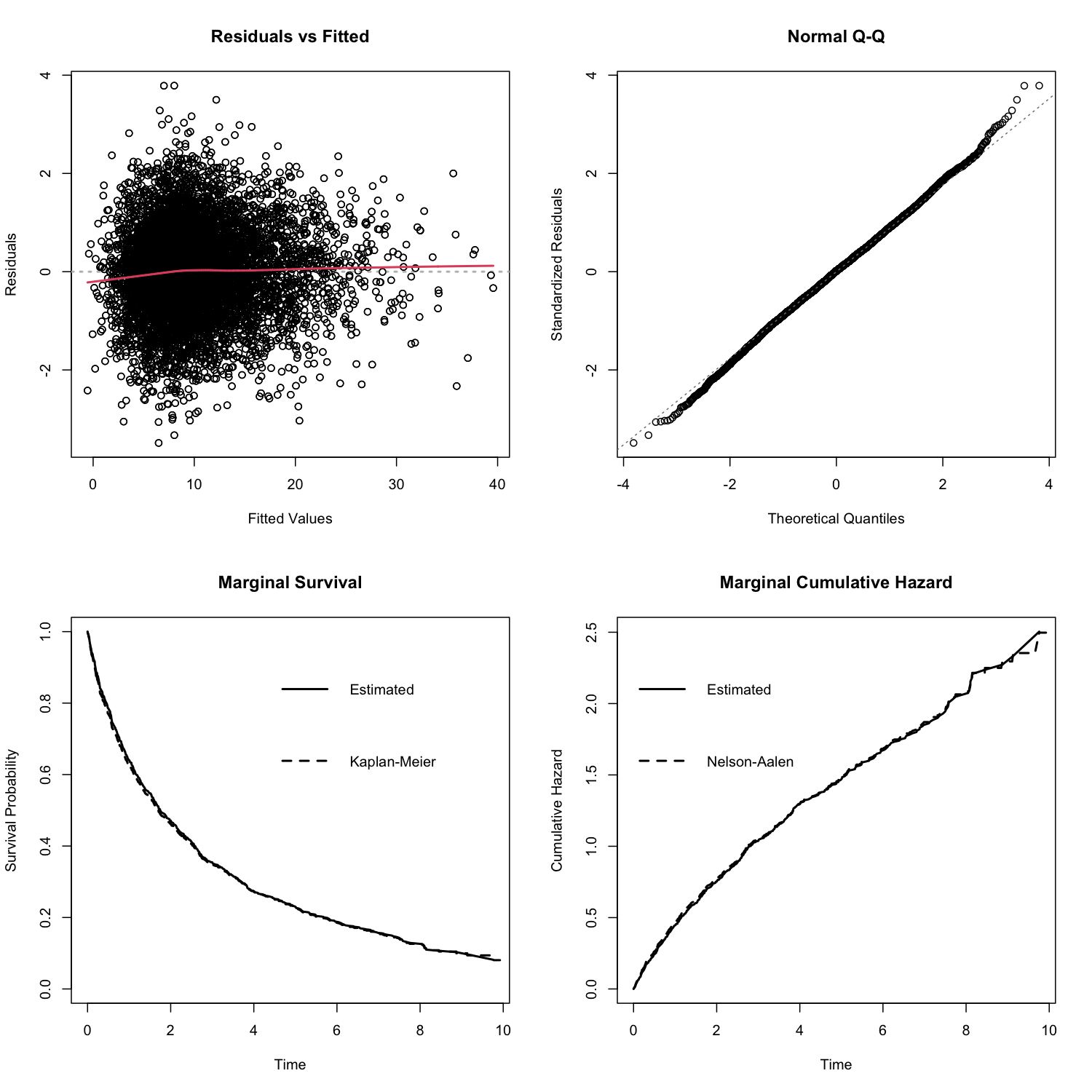}
    \caption{Standard diagnostic plots for \code{fit}. It displays four panels, including subject-specific residuals for the longitudinal process versus their corresponding fitted values (top-left), a normal Q-Q plot of the residuals of the standardized subject-specific residuals for the longitudinal process (top-right), an estimate of the marginal survival function for the event process (bottom-left), and an estimate of the marginal cumulative hazard function for the event process.}
    \label{fig:plotjmcs}
\end{figure}

\begin{figure}[!ht]
    \centering
    \includegraphics[width=0.7\linewidth]{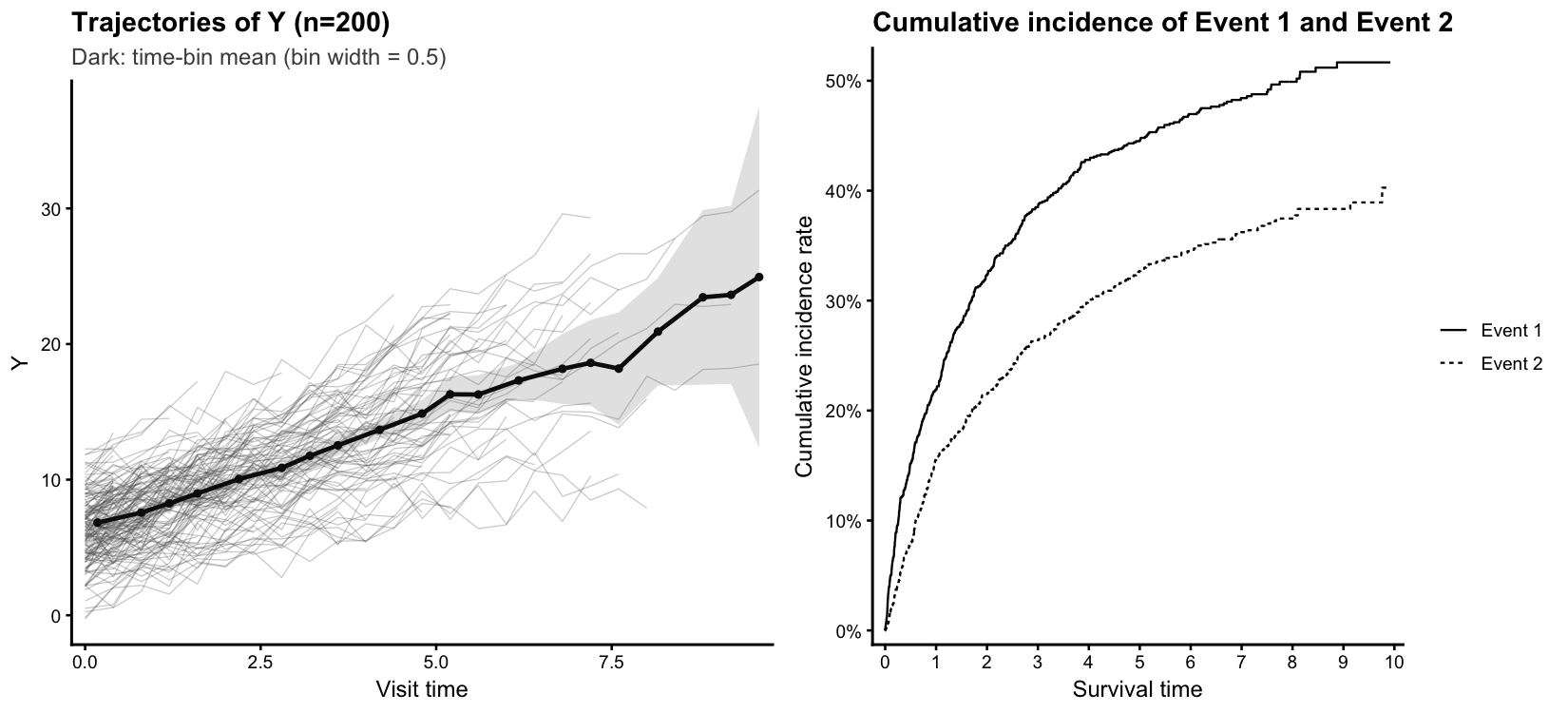}
    \caption{Exploratory diagnostic plots produced by \code{timeplot()} for \code{fit}. The left panel displays the empirical longitudinal biomarker mean trajectory together with raw longitudinal profiles from 200 randomly selected subjects. The right panel displays the empirical cumulative incidence rates for the primary and competing events.}
    \label{fig:plotjmcs2}
\end{figure}

\begin{figure}[!ht]
    \centering
    \includegraphics[width=0.7\linewidth]{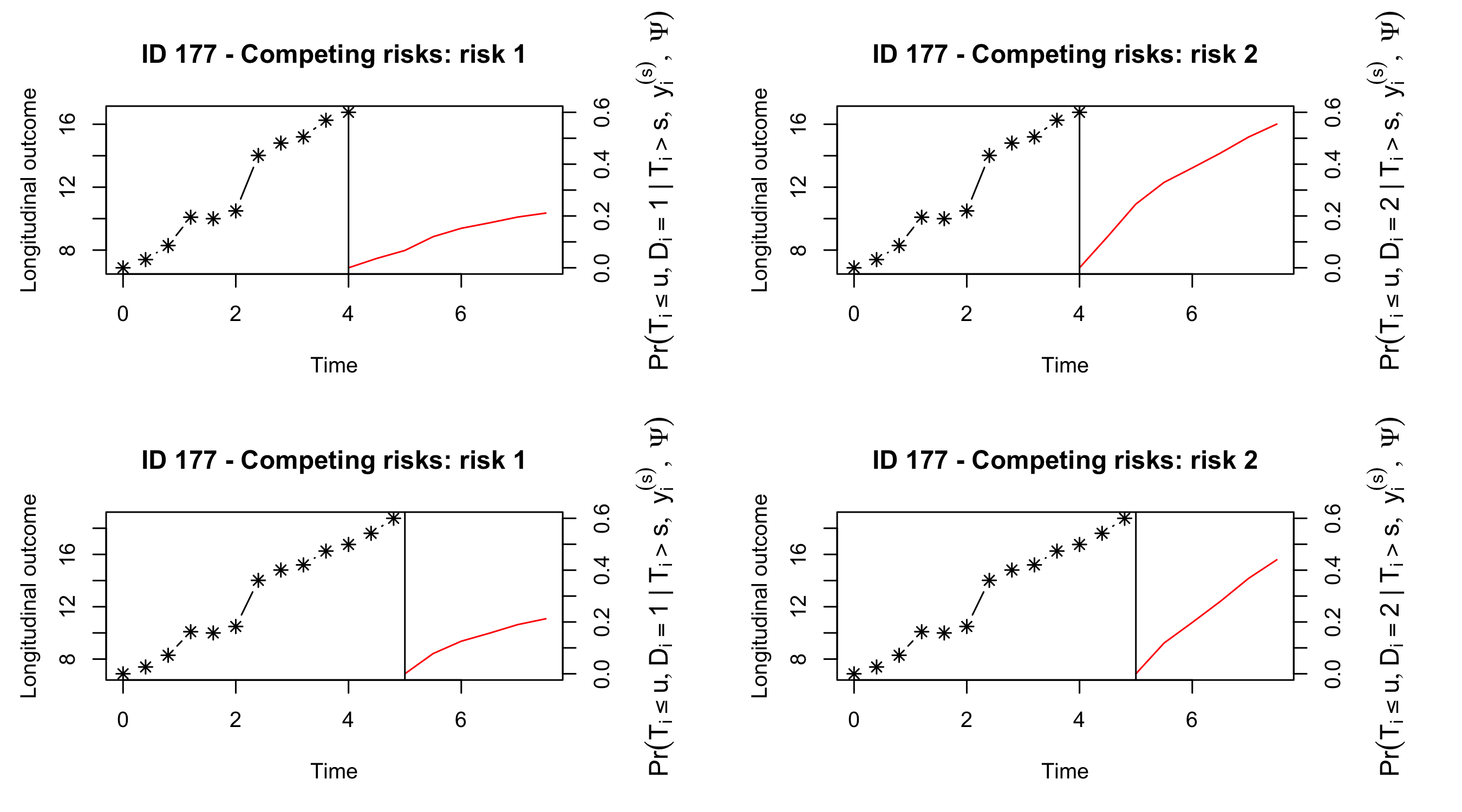}
    \caption{Predicted cumulative incidence functions of two risks for ID 117 who is event-free at pre-specified landmark times: $s=4$ (top panels) and $s=5$ (bottom panels). The asterisks inside each panel show the raw longitudinal measures by the end of the landmark time. The red lines depict the estimated probabilities of $P_{i^*k}(u,s)$. The black vertical lines mark the landmark time for prediction.}
    \label{fig:plotjmcs3}
\end{figure}

An additional exploratory diagnostic visualization can be obtained using
\code{timeplot()}, which summarizes the observed longitudinal trajectories and the event process. Figure~\ref{fig:plotjmcs2} displays the empirical longitudinal biomarker mean trajectory together with raw longitudinal profiles from 200 randomly selected subjects and the empirical cumulative incidence rates for the primary and competing events. Here, \code{time_bin_width = 0.5} specifies the width of the derived follow-up time bins used for calculating the raw mean repeated measures in the left panel, and \code{n.obs = 200} controls the number of randomly selected subject trajectories displayed. The arguments \code{fail_code} and \code{cr_code} identify the primary and competing event types used in the right panel.
\begin{Code}
R> timeplot(
+    object = fit,
+    biomarker = Y,
+    id_col = ID,
+    time_col = time,
+    time_bin_width = 0.5,
+    fail_code = 1,
+    cr_code = 2,
+    censor_code = 0,
+    primary_event_label = "Event 1",
+    competing_event_label = "Event 2",
+    x_lab = "Visit time",
+    event_x_lab = "Survival time",
+    n.obs = 200
+  )
\end{Code}
\paragraph{Dynamic prediction}
Finally, we illustrate the calculation of subject-specific cumulative incidence
probabilities for competing-risks outcomes. The dynamic prediction procedure
described in Section~\ref{sec:DP} is implemented in \code{survfitJM()}, which
takes a fitted joint model object, longitudinal measurements for new subjects,
and their corresponding baseline covariate and event-time information as input.
In this example, we select one subject, and use the longitudinal history observed up to the landmark time $s$ specified by \code{last.time}. The function \code{survfitJM()} estimates the cumulative incidence probabilities
\(P_{i^*k}(u, s \mid \widehat{\Psi})\) for the horizon times
\(u\), given that this subject is event-free at the landmark time $s$. We illustrate dynamic prediction by considering two landmark times, \code{Last.time = 4} and \code{Last.time = 5}. For each landmark time, cumulative incidence probabilities are evaluated at the user-specified horizon times $u$ through the argument \code{u}. The option \code{method = "GH"} requests Gauss–Hermite quadrature for numerical integration over the random-effects distribution. The argument \code{Last.time} is set \code{NULL} as default, the subject's observed event time. Alternatively, users may specify a numeric landmark time, as shown below.
\begin{Code}
R> Last.time <- 4
R> ND <- ydata[ydata$ID == 177 & ydata$time <= Last.time, ]
R> ID <- unique(ND$ID)
R> NDc <- cdata[cdata$ID %in% ID, ]
R> survfit <- survfitJM(
+    fit,
+    ynewdata = ND,
+    cnewdata = NDc,
+    Last.time = Last.time,
+    u = seq(4.5, 7.5, by = 0.5),
+    method = "GH",
+    obs.time = "time"
+  )
R> survfit
\end{Code}
\begin{Verbatim}
Prediction of Conditional Probabilities of Event
based on the pseudo-adaptive Gauss-Hermite quadrature rule with 3 quadrature points
$`177`
  times       CIF1      CIF2
1   4.0 0.00000000 0.0000000
2   4.5 0.03610781 0.1207161
3   5.0 0.06698280 0.2460498
4   5.5 0.11996821 0.3296914
5   6.0 0.15245170 0.3854109
6   6.5 0.17342389 0.4427040
7   7.0 0.19579361 0.5047166
8   7.5 0.21115206 0.5547216    
\end{Verbatim}
\begin{Code}
R> Last.time <- 5
R> ND <- ydata[ydata$ID == 177 & ydata$time <= Last.time, ]
R> survfit2 <- survfitJM(
+    fit,
+    ynewdata = ND,
+    cnewdata = NDc,
+    Last.time = Last.time,
+    u = seq(5.5, 7.5, by = 0.5),
+    method = "GH",
+    obs.time = "time"
+  )
R> survfit2
\end{Code}
\begin{Verbatim}
Prediction of Conditional Probabilities of Event
based on the pseudo-adaptive Gauss-Hermite quadrature rule
with 3 quadrature points

$`177`
  times       CIF1      CIF2
1   5.0 0.00000000 0.0000000
2   5.5 0.07793918 0.1187649
3   6.0 0.12577458 0.1980484
4   6.5 0.15669665 0.2797480
5   7.0 0.18970961 0.3684050
6   7.5 0.21241293 0.4401168
\end{Verbatim}
The resulting object contains subject-specific cumulative incidence predictions
for each event type across the requested prediction horizons. The estimated prediction
trajectories can also be visualized using the corresponding \code{plot()}
method for objects returned by \code{survfitJM()}, as displayed in Figure~\ref{fig:plotjmcs3}. The \code{ylim.surv} argument controls the y-axis range of the predicted survival probability or cumulative incidence rate.
\begin{Code}
R> oldpar <- par(mfrow = c(2, 2), mar = c(5.1, 4.1, 4.1, 6.5))
R> plot(survfit, include.y = TRUE, ylim.surv = c(0, 0.6))
R> plot(survfit2, include.y = TRUE, ylim.surv = c(0, 0.6))

\end{Code}

\paragraph{Prediction assessment}
Finally, we evaluate the accuracy of dynamic predictions using a generic function
\code{DynPredAcc()}, which implements cross-validated measures of prediction performance for competing risks data. Given a fitted joint model, the function repeatedly constructs training and validation sets, obtains subject-specific cumulative incidence predictions at pre-specified landmark and horizon times, and summarizes predictive accuracy using measures such as time-dependent AUC, C-index,
Brier score, and MAEQ.
\begin{Code}
R> res <- DynPredAcc(
+    object = fit,
+    landmark.time = 5,
+    horizon.time = c(6, 6.5, 7),
+    obs.time = "time",
+    method = "GH",
+    n.cv = 3,
+    metrics = c("AUC", "Cindex", "Brier Score", "MAEQ")
+  )
\end{Code}
Here, \code{landmark.time = 5} specifies the landmark time at which subjects who remain event-free are selected for prediction in the validation set, whereas \code{horizon.time = c(6, 6.5, 7)} defines the future horizon times used for evaluating dynamic prediction accuracy. The argument \code{n.cv = 3} requests three-fold cross-validation, and \code{metrics} specifies the prediction accuracy measures to be computed, including the
time-dependent AUC, C-index, Brier score, and MAEQ. The option
\code{method = "GH"} specifies Gauss--Hermite quadrature for numerical integration over the random effects distribution. Finally, the \code{summary()} method is used to display the resulting prediction accuracy measures across the specified horizon times for both events.
\begin{Code}
R> summary(res, metric = "AUC")

Expected AUC at the landmark time of 5 
based on 3 fold cross validation
  Horizon Time   AUC1   AUC2
1          6.0 0.7315 0.6991
2          6.5 0.7009 0.7455
3          7.0 0.6940 0.7743
R> summary(res, metric = "Cindex")

Expected Cindex at the landmark time of 5 
based on 3 fold cross validation
  Horizon Time Cindex1 Cindex2
1          6.0  0.6989  0.6863
2          6.5  0.6977  0.6876
3          7.0  0.6955  0.6915
R> summary(res, metric = "Brier Score")

Expected Brier Score at the landmark time of 5 
based on 3 fold cross validation
  Horizon Time Brier Score 1 Brier Score 2
1          6.0        0.0902        0.0695
2          6.5        0.1122        0.0935
3          7.0        0.1382        0.1264
R> summary(res, metric = "MAEQ")

Mean absolute error across quantiles of predicted risk scores at the 
landmark time of 5 
based on 3 fold cross validation
  Horizon Time  MAEQ1  MAEQ2
1          6.0 0.0832 0.0532
2          6.5 0.0925 0.0538
3          7.0 0.1090 0.0944
\end{Code}
In addition to these time-dependent prediction accuracy measures, here we illustrate the use of \code{Concordance()} with the fitted object \code{fit} returned by \code{jmcs()}. The argument \code{n.cv = 3} specifies three-fold cross-validation, while \code{seed = 100} ensures reproducibility of the
cross-validation partition.
\begin{Code}
R> Con <- Concordance(seed = 100, object = fit, n.cv = 3)
R> summary(Con)
\end{Code}
\begin{Code}
  Concordance1 Concordance2
1       0.8327       0.7973
\end{Code}

\subsection{Fitting a joint model with multiple longitudinal biomarkers}
\label{sec:expmvjmcs}
We next illustrate the multivariate joint modeling framework implemented in
\code{mvjmcs()}. While \code{jmcs()} and \code{JMMLSM()} focus on a single
longitudinal biomarker, \code{mvjmcs()} allows multiple longitudinal biomarkers
to be modeled jointly and associated with the competing-risks event process through shared random effects.

For illustration, we use the accompanying \code{simmvJMdata()} function to simulate a dataset containing three longitudinal biomarkers and competing-risks outcomes. Details of the arguments in \code{simmvJMdata()} and the corresponding generative joint model formulas are provided in Appendix \ref{appen:mvjmcs}. The resulting longitudinal dataset, \code{mvydata},
contains repeated measurements for three biomarkers; the competing-risks
dataset, \code{mvcdata}, contains the observed event times,
competing-risk indicators, and baseline covariates. 
\begin{Code}
R> mvdat <- simmvJMdata(...)
R> mvydata <- mvdat[["mvydata"]]
R> mvcdata <- mvdat[["mvcdata"]]
\end{Code}
\paragraph{Model fitting}
Here we fit a joint model for multiple longitudinal biomarkers using \code{mvjmcs()}. In particular, we fit the following tri-variate joint model:
\begin{eqnarray*}
Y_{i1}(t_{ij})
&=&
\beta_{10}
+\beta_{11}X_{i1}
+\beta_{12}X_{i2}
+b_{i10}
+\epsilon_{i1}(t_{ij}), \cr
Y_{i2}(t_{ij})
&=&
\beta_{20}
+\beta_{21}X_{i1}
+\beta_{22}X_{i2}
+\beta_{23}t_{ij}
+b_{i20}
+b_{i21}t_{ij}
+\epsilon_{i2}(t_{ij}), \cr
Y_{i3}(t_{ij})
&=&
\beta_{30}
+\beta_{31}X_{i1}
+\beta_{32}X_{i2}
+\beta_{33}t_{ij}
+\beta_{34}t_{ij}^2
+b_{i30}
+b_{i31}t_{ij}
+b_{i32}t_{ij}^2
+\epsilon_{i3}(t_{ij}),
\end{eqnarray*}
where
\begin{eqnarray*}
\epsilon_{ig}(t_{ij}) &\sim& N(0,\sigma_g^2), \quad g=1,2,3,
\end{eqnarray*}
and
\[
\lambda_{ik}(t \mid X_i,b_i)
=
\lambda_{0k}(t)
\exp\left\{
X_i^\top \gamma_k
+
\alpha_{1k} b_{i10}
+
\alpha_{2k}^\top
\begin{pmatrix}
b_{i20}\\
b_{i21}
\end{pmatrix}
+
\alpha_{3k}^\top
\begin{pmatrix}
b_{i30}\\
b_{i31}\\
b_{i32}
\end{pmatrix}
\right\},
\quad k=1,2.
\]
\begin{Code}
R> mvydata$timesq <- mvydata$time^2
R> fit.mv <- mvjmcs(
+    ydata = mvydata,
+    cdata = mvcdata,
+    long.formula = list(
+      Y1 ~ X1 + X2,
+      Y2 ~ X1 + X2 + time,
+      Y3 ~ X1 + X2 + time + timesq
+    ),
+    random = list(
+      ~ 1 | ID,
+      ~ time | ID,
+      ~ time + timesq | ID
+    ),
+    surv.formula = Surv(survtime, cmprsk) ~ X1 + X2,
+    control = mvjmcs_control(
+      cpu.cores = parallel::detectCores(),
+      opt = "optim"
+    )
+  )
\end{Code}
\begin{Code}
R> fit.mv

Call:
 mvjmcs(ydata = mvydata, cdata = mvcdata, long.formula = list(Y1 ~ X1 + X2, 
 Y2 ~ X1 + X2 + time, Y3 ~ X1 + X2 + time + timesq), random = list(~1 | ID, 
 ~time | ID, ~time + timesq | ID), surv.formula = Surv(survtime, cmprsk) ~ X1 + X2, 
 control = mvjmcs_control(cpu.cores = parallel::detectCores(), opt = "optim")) 

Data Summary:
Number of observations: 30399 
Number of groups: 5000 

Proportion of competing risks: 
Risk 1 : 37.68 %
Risk 2 : 23.56 %

Model Type: joint modeling of multivariate longitudinal continuous and 
competing risks data 

Model summary:
Runtime: 7.38 minutes 
Longitudinal process: linear mixed effects model
Event process: cause-specific Cox proportional hazard model with non-parametric 
baseline hazard

Fixed effects in the longitudinal sub-model:  list(Y1 ~ X1 + X2, 
Y2 ~ X1 + X2 + time, Y3 ~ X1 + X2 + time + timesq) 

                 Estimate     SE  Z value  p-val
(Intercept)_bio1   4.9137 0.0459 107.0396 0.0000
X1_bio1            1.6378 0.0709  23.0969 0.0000
X2_bio1            2.0161 0.0120 168.6426 0.0000
(Intercept)_bio2   9.6785 0.0620 156.0642 0.0000
X1_bio2            0.8624 0.0981   8.7915 0.0000
X2_bio2            1.9436 0.0165 118.1235 0.0000
time_bio2          0.9302 0.0205  45.3860 0.0000
(Intercept)_bio3   7.7424 0.0643 120.3354 0.0000
X1_bio3            1.3027 0.0945  13.7923 0.0000
X2_bio3            1.4942 0.0162  91.9554 0.0000
time_bio3          0.8180 0.0274  29.8007 0.0000
timesq_bio3        0.4041 0.0173  23.3305 0.0000


Residual error:
           Variance StdDev
sigma_bio1   0.9880 0.9940
sigma_bio2   0.9944 0.9972
sigma_bio3   0.9871 0.9935

Fixed effects in the survival sub-model:  Surv(survtime, cmprsk) ~ X1 + X2 

     Estimate     SE Z value  p-val
X1_1   0.7776 0.0810  9.5982 0.0000
X2_1   0.4223 0.0163 25.9050 0.0000
X1_2  -0.4510 0.0895 -5.0416 0.0000
X2_2   0.4544 0.0179 25.3830 0.0000

Association parameters:                 
                  Estimate     SE  Z value  p-val
(Intercept)_1bio1  -0.4705 0.0164 -28.7638 0.0000
(Intercept)_1bio2   0.4474 0.0135  33.1669 0.0000
time_1bio2          0.5973 0.0443  13.4906 0.0000
(Intercept)_1bio3   0.2829 0.0111  25.3772 0.0000
time_1bio3          0.3077 0.0517   5.9573 0.0000
timesq_1bio3        0.1283 0.0598   2.1457 0.0319
(Intercept)_2bio1   0.4590 0.0196  23.4743 0.0000
(Intercept)_2bio2   0.4599 0.0158  29.0535 0.0000
time_2bio2          0.6915 0.0554  12.4790 0.0000
(Intercept)_2bio3   0.2741 0.0126  21.7031 0.0000
time_2bio3          0.1715 0.0604   2.8424 0.0045
timesq_2bio3       -0.0821 0.0730  -1.1245 0.2608


Random effects:                 
  bio 1 :  ~1 | ID 
  bio 2 :  ~time | ID 
  bio 3 :  ~time + timesq | ID 
           StdDev   Intr1  Intr2   time2  Intr3  time3
Intercept1 2.2907                                     
Intercept2 3.1675 -0.0137                             
time2      1.0069 -0.0039 0.0354                      
Intercept3 3.1414 -0.0202 0.0187  0.0062              
time3      1.0060  0.0515 0.0356 -0.0011 0.0140       
timesq3    0.7255 -0.0301 0.0142 -0.0076 0.0375 0.0183
\end{Code}
The output is summarized into four main components: the longitudinal fixed effects, error variances, survival fixed effects, and association parameters. Parameters are labeled using descriptive suffixes corresponding to each biomarker (\code{bio1}, \code{bio2}, and \code{bio3}). In this example, the three biomarkers are modeled using different longitudinal submodels, illustrating the flexibility of \code{mvjmcs()} to accommodate biomarker-specific fixed-effects and random-effects structures. The matrix printed at the end of the output summarizes the standard deviations and correlations among all random effects across biomarkers. The dimension of the covariance matrix depends on the random-effects specification for each biomarker. In this example, biomarker 1 includes a random intercept, biomarker 2 includes a random intercept and slope, and biomarker 3 includes a random intercept together with random slope and random quadratic time effects.

In addition, the model summary reports the total runtime required for model fitting. In this example, parallel computing was enabled through \code{cpu.cores = parallel::detectCores()}, allowing \code{mvjmcs()} to utilize all available CPU cores during model fitting. The model was fitted to 5000 subjects and 30399 longitudinal observations in approximately 7.38 minutes on the computing platform used for this analysis.

\begin{figure}[!ht]
    \centering
    \includegraphics[width=0.8\linewidth]{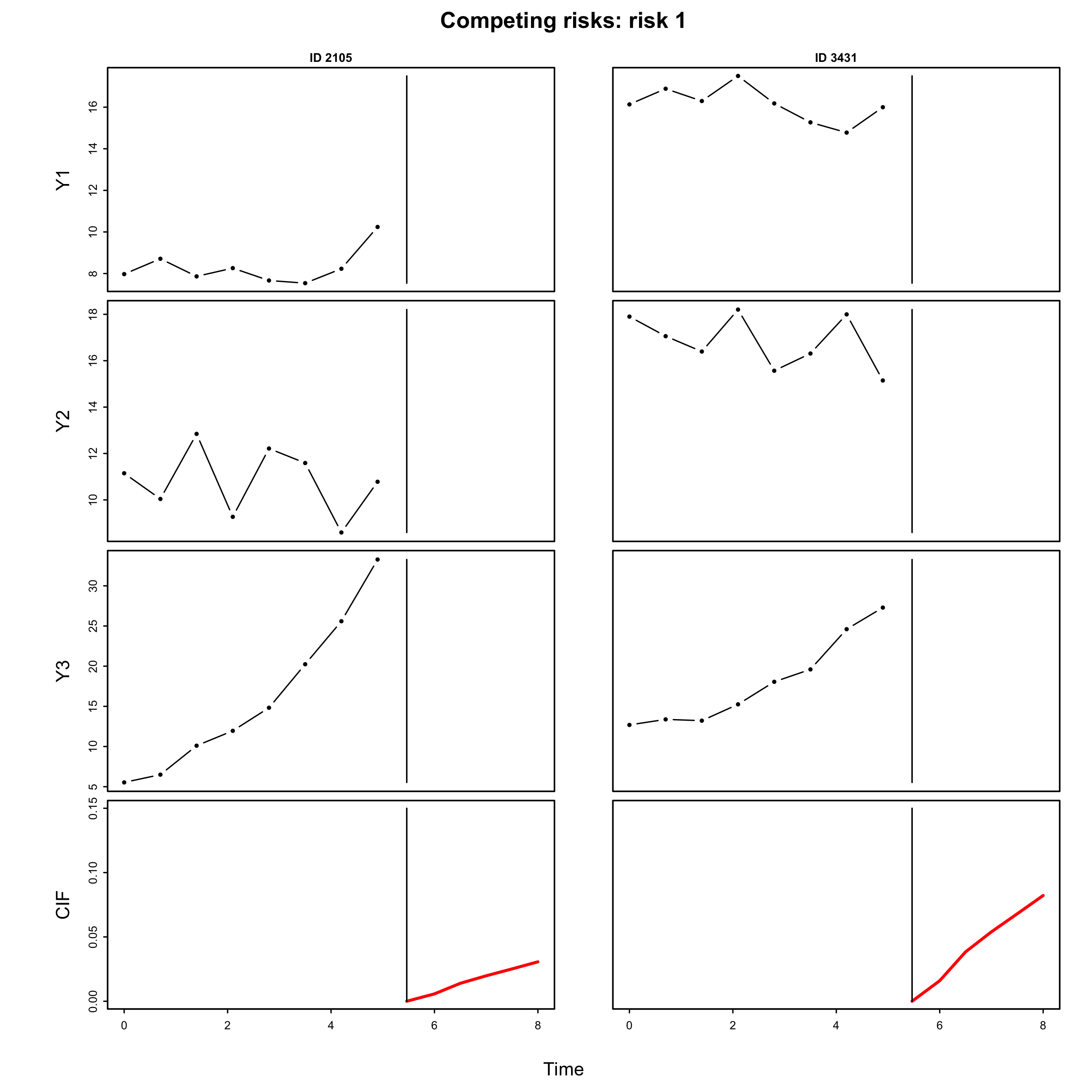}
    \caption{Predicted cumulative incidence functions of failure type 1 for ID 2105 and 3431 who is event-free by the end of the observation. The asterisks inside each panel show the raw longitudinal measures by the end of observation. The red lines depict the estimated probabilities of $P_{i^*k}(u,s)$. The black vertical lines mark the landmark time (observed event time) for prediction.}
    \label{fig:plotmvjmcs}
\end{figure}

\paragraph{Dynamic prediction}
Likewise in Section~\ref{sec:expjmcs}, we select two subjects and again use \code{survfitJM()} to estimate the cumulative incidence probabilities:
\begin{Code}
R> ND <- mvydata[mvydata[["ID"]] %in% c(2105, 3431), ]
R> ID <- unique(ND[["ID"]])
R> NDc <- mvcdata[mvcdata[["ID"]] %in% ID, ]
R> survfit <- survfitJM(
+    object = fit.mv,
+    ynewdata = ND,
+    cnewdata = NDc,
+    Last.time = NULL,
+    u = seq(6, 8, by = 0.5),
+    obs.time = "time"
+  )
R> survfit
\end{Code}
\begin{Code}
Prediction of Conditional Probabilities of Event
based on the first order approximation
$`2105`
     times        CIF1       CIF2
1 5.464087 0.000000000 0.00000000
2 6.000000 0.005737536 0.00978133
3 6.500000 0.013965795 0.01480255
4 7.000000 0.019792605 0.02059604
5 7.500000 0.025141000 0.02862488
6 8.000000 0.030642981 0.03688233

$`3431`
     times       CIF1       CIF2
1 5.465175 0.00000000 0.00000000
2 6.000000 0.01597835 0.02056702
3 6.500000 0.03840781 0.03090408
4 7.000000 0.05397257 0.04262663
5 7.500000 0.06801710 0.05854445
6 8.000000 0.08218845 0.07458967
\end{Code}
The \code{plot()} method also visualizes subject-specific dynamic predictions from a multivariate joint model. The method displays the observed longitudinal measurements for all biomarkers together with the corresponding predicted survival probability or cumulative incidence rate. The observed longitudinal measurements for all biomarkers are shown in vertically stacked panels, allowing users to examine the evolution of multiple biomarkers and their relationship with the predicted event risk over time (see Figure~\ref{fig:plotmvjmcs}). When \code{CompetingRisk = TRUE}, the \code{risk} argument specifies the failure type for which cumulative incidence probabilities are displayed. For example, the following command plots the dynamic prediction results for ID 2105 and 3431, which displays the stacked plot of observed biomarker trajectories of all three longitudinal outcomes and cumulative incidence function for failure type 1. Here, we specify \code{Last.time = NULL} to use their observed event times as the landmark time for survival prediction.
\begin{Code}
R> plot(survfit, risk = 1, ylim.surv = c(0, 0.15))
\end{Code}

\subsection{Fitting a joint model with heterogeneous WS variability}
\label{sec:expJMMLSM}
In this section, we illustrate the use of \code{JMMLSM()} for fitting a joint model that allows heterogeneous WS variability in the longitudinal process. Here, we use the accompanying \code{simJMWSVdata()} function to simulate the longitudinal dataset \code{ydatah} and the competing-risks dataset \code{cdatah}. Details of the arguments in \code{simJMWSVdata()} and the corresponding generative joint model formulas are provided in Appendix \ref{appen:JMMLSM}.
\begin{Code}
R> library("FastJM")
R> dat <- simJMWSVdata(...)
R> ydatah <- dat[["ydatah"]]
R> cdatah <- dat[["cdatah"]]
\end{Code}
\paragraph{Model fitting}
Before fitting the joint models with a single longitudinal biomarker with heterogeneous WS variability, we first fit a standard joint model using
\code{jmcs()}, which assumes a common residual variance for all subjects and
measurement occasions. This preliminary analysis illustrates how the
diagnostic functions \code{plot()} and \code{timeplot()} can be used to
identify potential inadequacy of the homogeneous WS variance assumption. The standard joint model and its implementation code is given by
\[
Y_i(t_{ij})
=
\beta_0
+\beta_1 X_{i1}
+\beta_2 X_{i2}
+\beta_3 X_{i3}
+\beta_4 t_{ij}
+b_{i0}
+b_{i1} t_{ij}
+\varepsilon_i(t_{ij}), \quad
\varepsilon_i(t_{ij})
\sim
N\!\left(0,\sigma^2\right),
\]
\[
\lambda_{ik}(t \mid X_i,b_i,w_i)
= \lambda_{0k}(t) \exp\!\left(X_i^\top\gamma_k + \alpha_k^\top b_i\right), \quad k=1,2.
\]
\begin{Code}
R> fit.homo <- jmcs(
+    long.formula = Y ~ X1 + X2 + X3 + time,
+    surv.formula = Surv(survtime, cmprsk) ~ X1 + X2 + X3,
+    random = ~ time | ID,
+    ydata = ydatah,
+    cdata = cdatah,
+    control = jmcs_control(quadpoint = 3)
+  )
R> plot(fit.homo)
R> timeplot(
+    fit.homo,
+    biomarker = Y,
+    id_col = ID,
+    time_col = time,
+    time_bin_width = 0.5,
+    fail_code = 1,
+    cr_code = 2,
+    censor_code = 0,
+    primary_event_label = "Event 1",
+    competing_event_label = "Event 2",
+    x_lab = "Visit time",
+    event_x_lab = "Survival time",
+    n.obs = 200
+  )
\end{Code}
\begin{figure}[!ht]
    \centering
    \includegraphics[width=0.6\linewidth]{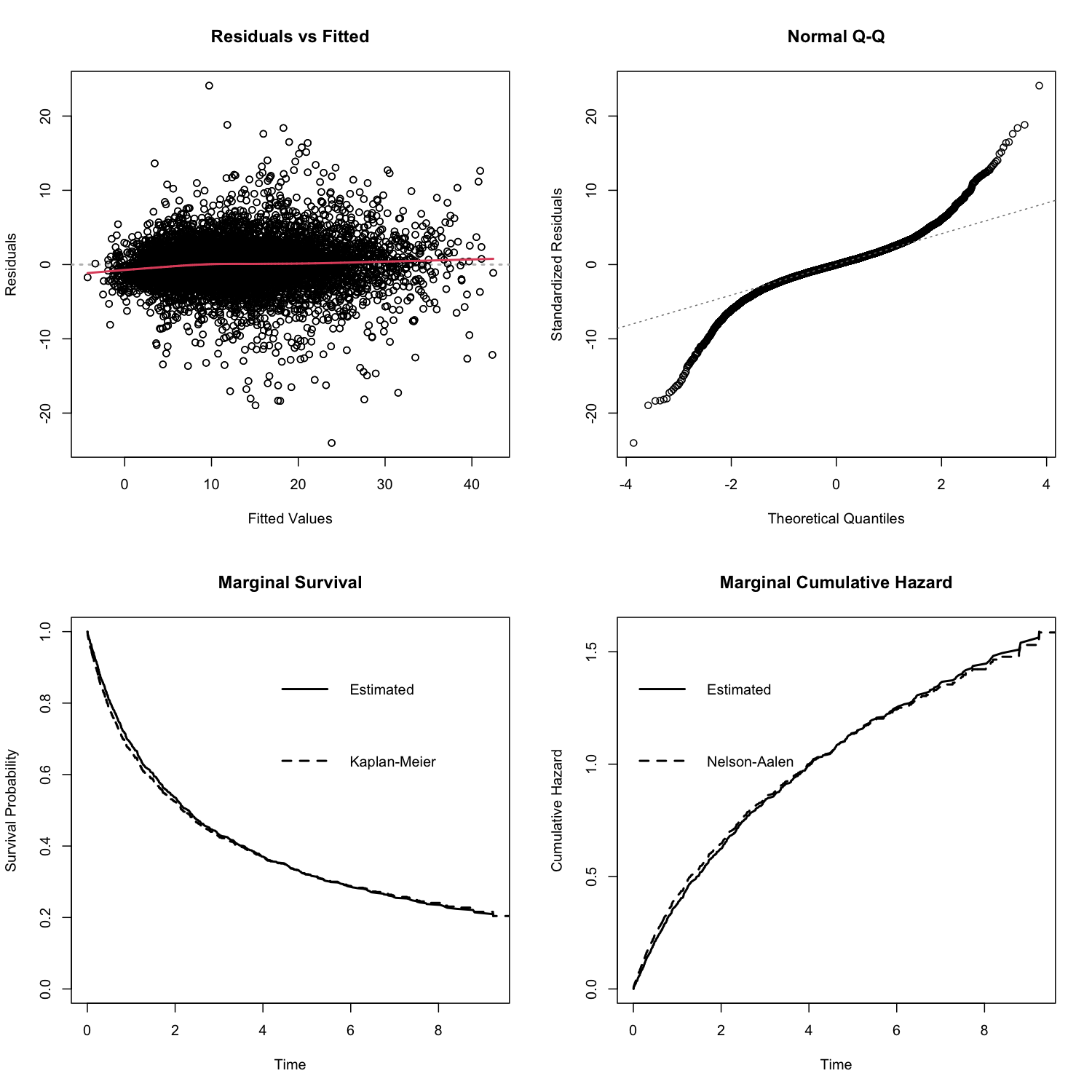}
    \caption{Standard diagnostic plots for \code{fit.homo}. It displays four panels, including subject-specific residuals for the longitudinal process versus their corresponding fitted values (top-left), a normal Q-Q plot of the residuals of the standardized subject-specific residuals for the longitudinal process (top-right), an estimate of the marginal survival function for the event process (bottom-left), and an estimate of the marginal cumulative hazard function for the event process.}
    \label{fig:plotJMMLSM1}
\end{figure}
\begin{figure}[!ht]
    \centering
    \includegraphics[width=0.7\linewidth]{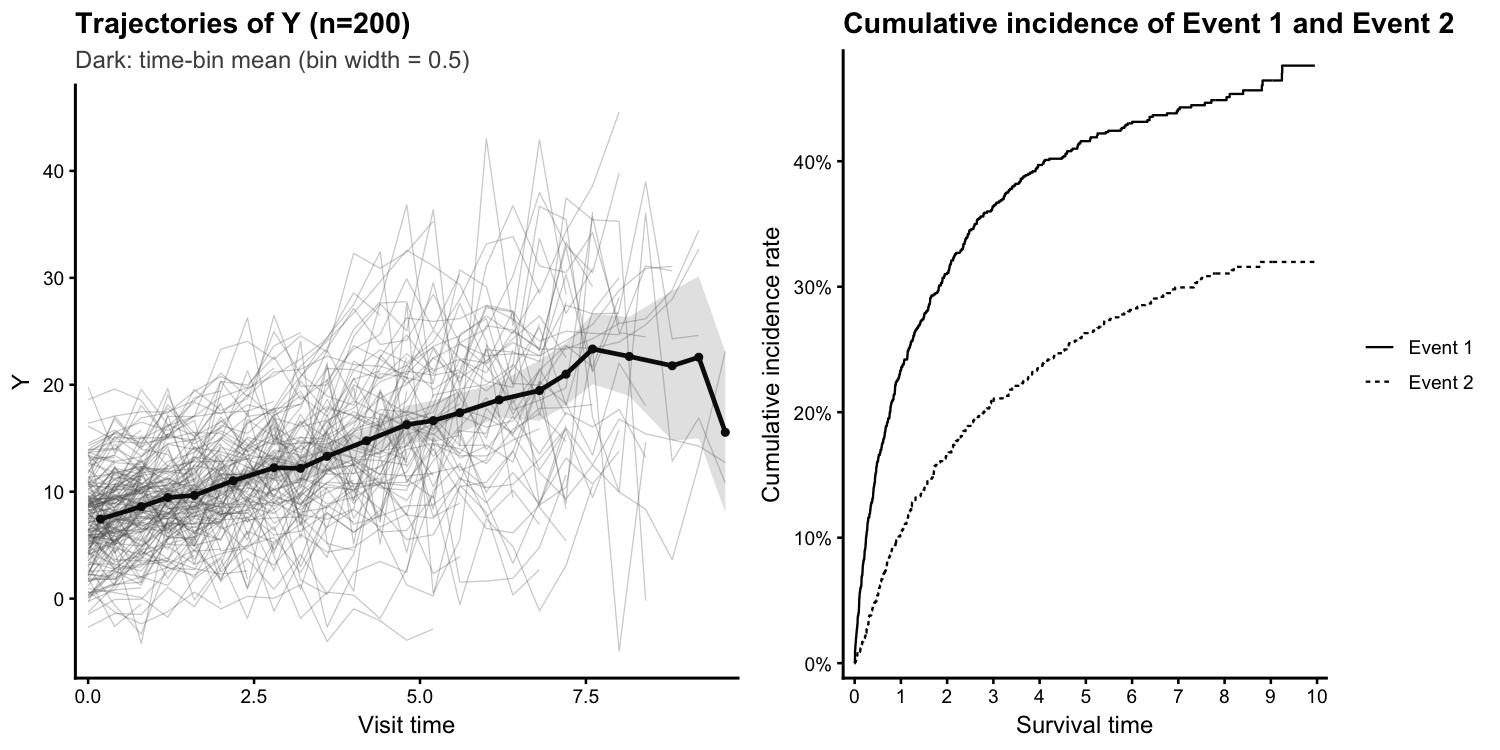}
    \caption{Exploratory diagnostic plots produced by \code{timeplot()} for \code{fit.homo}. The left panel displays the empirical longitudinal biomarker mean trajectory together with raw longitudinal profiles from 200 randomly selected subjects. The right panel displays the empirical cumulative incidence rates for the primary and competing events.}
    \label{fig:plotJMMLSM2}
\end{figure}

\paragraph{Model diagnostics}
Figure \ref{fig:plotJMMLSM1} produced by \code{plot()} shows that the dispersion of the residuals varies across the range of fitted values, and the normal Q--Q plot exhibits substantial departures from normality in both tails. In contrast, the model-based marginal survival and cumulative hazard curves are generally close to the corresponding Kaplan--Meier and Nelson--Aalen estimates. Figure \ref{fig:plotJMMLSM2} produced by \code{timeplot()} also shows considerable heterogeneity of WS variability, with the dispersion tending to increase over follow-up time. These patterns suggest that the assumption of a common residual variance may be inadequate. Thus, the most apparent model inadequacy arises from the longitudinal submodel.

These diagnostic findings motivate a joint model that explicitly allows heterogeneous WS variability. We therefore fit a joint model using \code{JMMLSM()} by incorporating the variance submodel~\eqref{eq3.2} with an additional random effect
\(\omega_i\). This allows the error variance to vary across subjects and
enables examination of the association between WS variability and event risk. In particular, we fit the following joint model:
\[
Y_i(t_{ij})
=
\beta_0
+\beta_1 X_{i1}
+\beta_2 X_{i2}
+\beta_3 X_{i3}
+\beta_4 t_{ij}
+b_{i0}
+b_{i1} t_{ij}
+\varepsilon_i(t_{ij}), \quad
\varepsilon_i(t_{ij})
\sim
N\!\left(0,\sigma_i^2(t_{ij})\right),
\]
\[
\log\{\sigma_i^2(t_{ij})\}
=
\tau_0
+\tau_1 X_{i1}
+\tau_2 X_{i2}
+\tau_3 X_{i3}
+\tau_4 t_{ij}
+\omega_i,
\]
\[
\lambda_{ik}(t \mid X_i,b_i,w_i)
= \lambda_{0k}(t) \exp\!\left(X_i^\top\gamma_k + \alpha_k^\top b_i + \nu_k \omega_i\right), \quad k=1,2.
\]
\begin{Code}
R> fit.ws <- JMMLSM(
+    long.formula = Y ~ X1 + X2 + X3 + time,
+    variance.formula = ~ X1 + X2 + X3 + time,
+    surv.formula = Surv(survtime, cmprsk) ~ X1 + X2 + X3,
+    random = ~ time | ID,
+    ydata = ydatah,
+    cdata = cdatah,
+    control = JMMLSM_control(quadpoint = 3)
+  )

R> fit.ws

Call:
 JMMLSM(cdata = cdatah, ydata = ydatah, long.formula = Y ~ X1 + X2 + X3 + time, 
 surv.formula = Surv(survtime, cmprsk) ~ X1 + X2 + X3, 
 variance.formula = ~X1 + X2 + X3 + time, random = ~time | ID, 
 control = JMMLSM_control(quadpoint = 3)) 

Data Summary:
Number of observations: 8803 
Number of groups: 1000 

Proportion of competing risks: 
Risk 1 : 44.8 %
Risk 2 : 30.2 %

Numerical intergration:
Method: adaptive Guass-Hermite quadrature
Number of quadrature points:  3 

Model Type: joint modeling of longitudinal continuous and competing risks data 
with the presence of intra-individual variability 

Model summary:
Longitudinal process: Mixed effects location scale model
Event process: cause-specific Cox proportional hazard model with non-parametric 
baseline hazard

Loglikelihood:  -26830.09 

Fixed effects in mean of longitudinal submodel:  Y ~ X1 + X2 + X3 + time 

            Estimate     SE Z value  p-val
(Intercept)   5.0804 0.1560 32.5695 0.0000
X1            1.7476 0.2162  8.0849 0.0000
X2            2.2265 0.1916 11.6195 0.0000
X3            0.9394 0.0551 17.0602 0.0000
time          1.9603 0.0514 38.1600 0.0000

Fixed effects in variance of longitudinal submodel:  log(sigma^2) ~ X1 + X2 + 
X3 + time 

            Estimate     SE Z value  p-val
(Intercept)   0.2984 0.0547  5.4585 0.0000
X1            0.5497 0.0702  7.8351 0.0000
X2           -0.0376 0.0606 -0.6207 0.5348
X3           -0.1683 0.0188 -8.9413 0.0000
time          0.3252 0.0081 40.0763 0.0000

Survival sub-model fixed effects:  Surv(survtime, cmprsk) ~ X1 + X2 + X3 

     Estimate     SE Z value  p-val
X1_1   0.7562 0.1383  5.4663 0.0000
X2_1   0.3225 0.1140  2.8280 0.0047
X3_1   0.5019 0.0390 12.8750 0.0000
                                  
X1_2 -0.5114 0.1404 -3.6421 0.0003
X2_2  0.5282 0.1159  4.5582 0.0000
X3_2  0.2426 0.0390  6.2138 0.0000

Association parameters:                 
                  Estimate     SE  Z value  p-val
(Intercept)_1      -0.3720 0.0259 -14.3677 0.0000
time_1              0.5921 0.1042   5.6814 0.0000
(Intercept)_2      -0.2141 0.0239  -8.9472 0.0000
time_2             -0.5184 0.1034  -5.0136 0.0000
var_(Intercept)_1   0.1676 0.1278   1.3121 0.1895
var_(Intercept)_2  -0.4819 0.1204  -4.0014 0.0001


Random effects:                 
  Formula: ~time | ID 
                StdDev  (Intr)    time
(Intercept)     3.2659                
time            0.9986 -0.0032        
var_(Intercept) 0.7592  0.0378 -0.0670
\end{Code}
The argument \code{variance.formula} specifies the fixed-effects covariates in the variance submodel~\eqref{eq3.2}, while
\code{random = ~ time | ID} specifies the random-effects structure for the
longitudinal mean submodel~\eqref{eq3.1}. Compared with \code{jmcs()},
\code{JMMLSM()} additionally estimates the fixed effects in the variance
submodel~\eqref{eq3.2} and the association between subject-specific WS
variability and the competing-risks event process. In addition to the fixed
effects in the longitudinal mean submodel~\eqref{eq3.1}, statistical inference
is also provided for the fixed effects in the variance submodel~\eqref{eq3.2}. The association parameters for \(\omega_i\), \code{var_(Intercept)_1} and
\code{var_(Intercept)_2}, quantify the effects of subject-specific WS
variability on the risks of the two competing events. Lastly, the estimated standard deviations and correlations among \(\omega_i\) (\code{var_(Intercept)}), \(b_{i0}\) (\code{(Intercept)}), and \(b_{i1}\) (\code{time}) are estimated and reported in the model output.

The post-estimation functions illustrated for \code{jmcs()} can also be
applied to the fitted model returned by \code{JMMLSM()}. In particular, \code{survfitJM()}
can be used to obtain the predicted subject-specific cumulative incidence rates that
incorporate both the longitudinal mean trajectory and the subject-specific WS
variability for estimating \(P_{i^*k}(u, s\mid \Psi)\), while \code{DynPredAcc()} can be used to evaluate dynamic prediction accuracy at pre-specified landmark and horizon times. Since their syntax and usage are the same as those presented for \code{jmcs()} in Section~\ref{sec:expjmcs}, we do not repeat the corresponding code here.

\section{Extended features: landmark multivariate joint modeling}
\label{sec:5}
The joint modeling methods described in the preceding sections adopt the shared random-effects parameterization, in which the longitudinal and event processes are linked through subject-specific random effects. This parameterization provides a flexible framework for characterizing the association between longitudinal biomarkers and competing-risks outcomes while enabling the scalable estimation algorithms implemented in \pkg{FastJM}. In particular, the computational complexity of these algorithms scales linearly with the sample size, making them well suited for large-scale biomedical applications such as electronic health records and biobank studies.

Despite these computational advantages, the shared random-effects parameterization does not directly accommodate several commonly used time-dependent latent association structures in the joint modeling literature, such as the current value and latent process parameterizations. These association structures relate the event risk to the underlying latent biomarker trajectory and are particularly attractive for dynamic prediction because they provide an interpretable characterization of how disease progression influences future clinical outcomes. However, incorporating these time-dependent association structures into semiparametric joint models generally requires substantially greater computational effort and is not directly compatible with the scalable estimation framework described in Section~\ref{sec:2}.

Landmarking provides an alternative framework for dynamic prediction by restricting the analysis to subjects who remain event-free at a pre-specified landmark time and modeling the subsequent event process conditional on survival beyond that time \citep{van2007dynamic}.
 Compared with conventional joint modeling, landmarking is computationally attractive because standard survival models can be fitted separately or jointly across clinically meaningful landmark times \citep{rizopoulos2017dynamic,li2023comparison}. Standard landmarking approaches, however, typically rely on the most recent observed biomarker value or simple summaries of the observed longitudinal history, rather than accounting for measurement error or the underlying latent biomarker trajectories. As a result, prediction accuracy may deteriorate when longitudinal measurements are sparse, irregularly spaced, or subject to substantial measurement error \citep{rizopoulos2017dynamic,putter2022landmarking,li2023comparison}.

To combine the strengths of joint modeling and landmarking, we propose a novel landmark multivariate joint modeling framework, which is implemented as an extended feature of \pkg{FastJM}. The proposed approach fits a multivariate joint model using subjects who remain event-free at a pre-specified landmark time while defining the post-landmark association between the longitudinal and event processes through the latent biomarker trajectories evaluated at the landmark time. Consequently, the proposed framework preserves the measurement-error correction and subject-specific trajectory estimation provided by joint modeling while introducing the flexibility of landmark-based dynamic prediction. At the same time, it retains the computational scalability of the estimation algorithms developed for the multivariate joint model. Additional methodological and computational details are provided in Appendix~\ref{Appen:supp_landmark_joint}.

For subjects who remain event-free at a pre-specified landmark time $s$, the longitudinal submodels are identical to those introduced in Section~\ref{sec:JMclass} and are given by
\[
Y_{ig}(t)=m_{ig}(t)+\sigma_g\epsilon_{ig}(t)
=X_{ig}^{\top}(t)\beta_g+Z_{ig}^{\top}(t)b_{ig}
+\sigma_g\epsilon_{ig}(t),
\qquad
\epsilon_{ig}(t)\stackrel{\text{i.i.d.}}{\sim}N(0,1),
\]
with
\[
b_i=(b_{i1}^{\top},\ldots,b_{iG}^{\top})^{\top}\sim N(0,\Sigma),
\]
where $m_{ig}(t)$ denotes the latent mean trajectory of biomarker $g$ for subject $i$. Unlike the standard multivariate joint model, the proposed landmark framework conditions on the landmark risk set $\{T_i>s\}$ and models the post-landmark event process through latent biomarker characteristics evaluated at the landmark time $s$. Specifically, for $t>s$, the cause-specific hazard for failure type $k$ is specified as
\begin{equation}
\lambda_{ik}(t\mid T_i>s,W_i,b_i)
=
\lambda_{0k}(t)
\exp\left\{
W_i^{\top}\gamma_k
+
\sum_{g=1}^{G}\alpha_{gk}A_{ig}(s)
\right\},
\qquad t>s,
\label{eq:landmark_hazard}
\end{equation}
where $\sum_{g=1}^{G}\alpha_{gk}A_{ig}(s)$ is the latent association function linking the longitudinal biomarkers to the post-landmark cause-specific hazard. 

The aboved landmark multivariate joint modeling framework accommodates different latent association structures through the specification of the association function $A_{ig}(s)$. This flexibility allows investigators to select the latent biomarker characteristic that is most appropriate for the scientific question of interest. The current implementation in \code{mvjmcs()} supports two commonly used parameterizations, corresponding to \code{latAsso = "present"} and \code{latAsso = "presentlp"}.

For \code{latAsso = "present"}, the association is defined by the subject-specific latent mean trajectory evaluated at the landmark time $s$, corresponding to the current value parameterization,
\begin{equation}
A_{ig}(s)=m_{ig}(s).
\label{eq:present}
\end{equation}

For \code{latAsso = "presentlp"}, the association is defined by the subject-specific random-effects contribution to the latent trajectory evaluated at the landmark time $s$, corresponding to the current value of the latent process parameterization,
\begin{equation}
A_{ig}(s)=Z_{ig}^{\top}(s)b_{ig}.
\label{eq:presentlp}
\end{equation}

Landmark fitting is requested with  \code{landmark = TRUE} and specifying a landmark time through the argument \code{s}. When landmark modeling is requested, the time variable used in the longitudinal submodels must be specified through \code{ytime}, and all longitudinal submodels must explicitly include this variable. Subjects who experience an event before the landmark time are automatically excluded from the analysis, and the event process is modeled conditional on survival beyond the landmark time $s$.

The following example fits a landmark multivariate joint model at a landmark time of five years using the data described in Section~\ref{sec:expmvjmcs}. For subjects who remain event-free at the landmark time $(s=5)$, the three longitudinal outcomes are modeled as
\[
\begin{aligned}
Y_{i1}(t_{ij})
={}&
\beta_{10}
+\beta_{11}X_{i1}
+\beta_{12}X_{i2}
+\beta_{13}t_{ij}
+b_{i10}
+\epsilon_{i1}(t_{ij}),
\\
Y_{i2}(t_{ij})
={}&
\beta_{20}
+\beta_{21}X_{i1}
+\beta_{22}X_{i2}
+\beta_{23}t_{ij}
+b_{i20}
+b_{i21}t_{ij}
+\epsilon_{i2}(t_{ij}),
\\
Y_{i3}(t_{ij})
={}&
\beta_{30}
+\beta_{31}X_{i1}
+\beta_{32}X_{i2}
+\beta_{33}t_{ij}
+\beta_{34}t_{ij}^{2}
+b_{i30}
+b_{i31}t_{ij}
+b_{i32}t_{ij}^{2}
+\epsilon_{i3}(t_{ij}).
\end{aligned}
\]
The measurement errors satisfy
\[
\epsilon_{ig}(t_{ij})
\sim N(0,\sigma_g^2),
\qquad g=1,2,3,
\]
and the subject-specific random effects are jointly distributed as
\[
b_i
=
\left(
b_{i10},
b_{i20},
b_{i21},
b_{i30},
b_{i31},
b_{i32}
\right)^\top
\sim N(0,\Sigma).
\]

The cause-specific hazard is modeled as 
\[
\lambda_{ik}(t\mid X_i,b_i,s)
=
\lambda_{0k}(t)
\exp\left\{
X_i^\top\gamma_k
+
\sum_{g=1}^{3}\alpha_{gk}\ell_{ig}(s)
\right\},
\qquad t>s,\quad k=1,2,
\]
where
\[
\begin{aligned}
\ell_{i1}(s)
&= b_{i10},\\
\ell_{i2}(s)
&= b_{i20}+b_{i21}s,\\
\ell_{i3}(s)
&= b_{i30}+b_{i31}s+b_{i32}s^2,
\end{aligned}
\]
with the latent association structure specified by \code{latAsso = "presentlp"}.

The following call to \code{mvjmcs()} implements this joint model. The three elements of \code{long.formula} define the fixed-effects components of the longitudinal submodels for \(Y_{i1}\), \(Y_{i2}\), and \(Y_{i3}\), respectively. The corresponding elements of \code{random} specify a random intercept for the first biomarker; a random intercept and linear time slope for the second biomarker; and a random intercept, linear time slope, and quadratic time effect for the third biomarker. The argument \code{surv.formula} includes \(X_{i1}\) and \(X_{i2}\) as baseline covariates in each cause-specific hazard model. Setting \code{landmark = TRUE} and \code{s = 5} restricts model fitting to subjects who remain event-free at five years. The argument \code{latAsso = "presentlp"} links the cause-specific hazards to the biomarker-specific latent random-effects contributions \(\ell_{ig}(s)\), while \code{ytime = "time"} identifies the longitudinal observation-time variable used to evaluate these contributions.
\begin{Code}
R> fit.mvlm <- mvjmcs(
+   mvydata, mvcdata,
+   long.formula = list(
+     Y1 ~ X1 + X2 + time,
+     Y2 ~ X1 + X2 + time,
+     Y3 ~ X1 + X2 + time + timesq
+   ),
+   random = list(
+     ~ 1 | ID,
+     ~ time | ID,
+     ~ time + timesq | ID
+   ),
+   surv.formula = Surv(survtime, cmprsk) ~ X1 + X2,
+   control = mvjmcs_control(
+     opt = "optim",
+     cpu.cores = parallel::detectCores()
+   ),
+   latAsso = "presentlp",
+   landmark = TRUE,
+   s = 5,
+   ytime = "time"
+ )

R> fit.mvlm

Call:
 mvjmcs(ydata = mvydata, cdata = mvcdata, long.formula = list(Y1 ~ X1 + X2 + time, 
Y2 ~ X1 + X2 + time, Y3 ~ X1 + X2 + time + timesq), random = list(~1 | ID, 
~time | ID, ~time + timesq | ID), surv.formula = Surv(survtime, cmprsk) ~ X1 + X2, 
control = mvjmcs_control(opt = "optim", cpu.cores = parallel::detectCores()), 
latAsso = "presentlp", landmark = TRUE, s = 5, ytime = "time") 

Data Summary:
Number of observations: 23951 
Number of groups: 2191 

Proportion of competing risks: 
Risk 1 : 7.07 %
Risk 2 : 4.47 %

Model Type: joint modeling of multivariate longitudinal continuous and 
competing risks data 

Model summary:
Landmark analysis: Yes (s = 5)
Latent association: current value of the latent process
Runtime: 37.5 seconds 
Longitudinal process: linear mixed effects model
Event process: cause-specific Cox proportional hazard model with non-parametric 
baseline hazard

Fixed effects in the longitudinal sub-model:  list(Y1 ~ X1 + X2 + time, 
Y2 ~ X1 + X2 + time, Y3 ~ X1 + X2 + time + timesq) 

                 Estimate     SE  Z value  p-val
(Intercept)_bio1   4.9219 0.0682  72.1223 0.0000
X1_bio1            1.9803 0.0889  22.2656 0.0000
X2_bio1            2.0114 0.0182 110.3848 0.0000
time_bio1         -0.0031 0.0028  -1.0988 0.2719
(Intercept)_bio2   8.1118 0.0899  90.2223 0.0000
X1_bio2            0.7924 0.1171   6.7670 0.0000
X2_bio2            1.6630 0.0238  69.7532 0.0000
time_bio2          0.7781 0.0208  37.3460 0.0000
(Intercept)_bio3   6.7350 0.1012  66.5367 0.0000
X1_bio3            1.3266 0.1354   9.7972 0.0000
X2_bio3            1.3431 0.0279  48.1670 0.0000
time_bio3          0.7223 0.0244  29.5474 0.0000
timesq_bio3        0.3804 0.0157  24.1849 0.0000


Residual error:
           Variance StdDev
sigma_bio1   0.9870 0.9935
sigma_bio2   0.9941 0.9970
sigma_bio3   0.9908 0.9954

Fixed effects in the survival sub-model:  Surv(survtime, cmprsk) ~ X1 + X2 

     Estimate     SE Z value  p-val
X1_1   0.3210 0.1682  1.9078 0.0564
X2_1   0.2133 0.0322  6.6254 0.0000
X1_2  -0.1887 0.2094 -0.9009 0.3677
X2_2   0.2265 0.0398  5.6943 0.0000

Association parameters:                 
            Estimate     SE Z value  p-val
alpha1_bio1  -0.4246 0.0493 -8.6087 0.0000
alpha1_bio2   0.1222 0.0171  7.1422 0.0000
alpha1_bio3   0.0087 0.0044  1.9807 0.0476
alpha2_bio1   0.4042 0.0625  6.4694 0.0000
alpha2_bio2   0.1357 0.0245  5.5500 0.0000
alpha2_bio3   0.0092 0.0057  1.6159 0.1061


Random effects:                 
  bio 1 :  ~1 | ID 
  bio 2 :  ~time | ID 
  bio 3 :  ~time + timesq | ID 
           StdDev   Intr1   Intr2   time2   Intr3  time3
Intercept1 2.0365                                       
Intercept2 2.6730  0.0596                               
time2      0.9758  0.0121 -0.1631                       
Intercept3 3.0368 -0.0149 -0.2004 -0.0991               
time3      0.9985  0.0901 -0.0666 -0.0382 -0.0391       
timesq3    0.7252 -0.0041 -0.0085 -0.0244  0.0264 0.0133
\end{Code}
The fitted model is summarized by printing the \code{mvjmcs} object. Compared with the output from the standard multivariate joint model in Section \ref{sec:expmvjmcs}, the summary additionally reports the landmark analysis and the selected latent association structure. Under \code{latAsso = "presentlp"}, a scalar association parameter is estimated for each biomarker, quantifying the effect of its latent process evaluated at the landmark time on the cause-specific event risks. This extension is particularly useful when the goal is to evaluate the event risk among subjects who remain event-free up to a landmark time, since it allows for assessing the latent associations between the longitudinal processes and cause-specific event risks within the risk set defined at that landmark time.

\section{Concluding remarks}
\label{sec:6}
\pkg{FastJM} provides a unified frequentist workflow for three semiparametric joint-model classes:  a  single longitudinal biomarker, multiple correlated longitudinal biomarkers, and a single longitudinal biomarker with heterogeneous WS variability. The package combines estimation with extraction, visualization, dynamic prediction, and prediction assessment. 

A distinguishing feature of \pkg{FastJM} is its emphasis on computational scalability. Within a unified expectation--maximization framework, the package combines model-specific approximation methods for evaluating the E-step with customized linear-scan algorithms for updating the nonparametric baseline hazards and computing profile-likelihood standard errors. These computational strategies substantially reduce the computational burden while maintaining estimation accuracy, making semiparametric joint modeling feasible for modern large-scale biomedical studies, including electronic health records and population-based biobanks.

This paper also introduced a novel landmark multivariate joint modeling framework that extends the modeling capabilities of \pkg{FastJM} beyond the shared random-effects parameterization. By combining landmark analysis with multivariate joint modeling, the proposed framework accommodates commonly used time-dependent latent association structures while retaining the computational advantages of the underlying estimation algorithms. Together with the three core modeling functions, this extension broadens the range of dynamic prediction problems that can be addressed within a unified software environment.

Future development of \pkg{FastJM} will focus on extending the framework to additional classes of joint models, including recurrent-event outcomes and non-Gaussian longitudinal responses, while preserving the computational scalability and unified software framework described in this paper.

\bibliographystyle{apalike}
\bibliography{reference}

\newpage
\appendix

\section{Simulation settings for illustrating examples}
\subsection{Illustration of generating the single-biomarker data and competing risks data}
\label{appen:jmcs}
To illustrate \code{jmcs()} in Section \ref{sec:expjmcs}, we generated one longitudinal biomarker and two competing event types with \code{simJMdata()}.  
\begin{Code}
R> data <- simJMdata(
+     seed = 100,
+     N = 1000,
+     increment = 0.4,
+     beta = c(5, 1.5, 2, 1, 2),
+     sigma2 = 1,
+     gamma1 = c(1, 0.5, 0.5),
+     gamma2 = c(-0.5, 0.5, 0.25),
+     alpha1 = c(1, 0.7),
+     alpha2 = c(-1, -0.5),
+     lambda1 = 0.05,
+     lambda2 = 0.1,
+     CL = 5,
+     CU = 10,
+     covb = diag(rep(1, 2)),
+     CR = TRUE
+ )
R> ydata <- data[["ydata"]]
R> cdata <- data[["cdata"]]
\end{Code}
For subject \(i = 1,\ldots,n\), three baseline covariates $X_i = (X_{i1}, X_{i2}, X_{i3})^{\top}$ were generated:
\[
X_{i1} \sim \mathrm{Bernoulli}(0.5), \qquad
X_{i2} \sim \mathrm{Uniform}(-1,1), \qquad
X_{i3} \sim N(1, 2^2).
\]
The subject-specific random effects were generated as
\[
b_i = (b_{i0}, b_{i1})^\top \sim N(0, \Sigma),
\]
where the random effects \(b_{i0}\) and \(b_{i1}\) represent the
subject-specific deviation in the baseline biomarker value and the
subject-specific time slope, respectively. \(\Sigma\) is specified by the argument \code{covb}.

Conditional on covariates and random effects, longitudinal measurements were generated from the stated linear mixed-effects model at equally spaced times $t_{ij} = j \times \texttt{increment}$ from baseline through the observed event-or-censoring time,
\[
Y_i(t_{ij})
=
\beta_0
+ \beta_1 X_{i1}
+ \beta_2 X_{i2}
+ \beta_3 X_{i3}
+ \beta_4 t_{ij}
+ b_{i0}
+ b_{i1} t_{ij}
+ \epsilon_i(t_{ij}), \quad
\epsilon_i(t_{ij}) \sim N(0, \sigma^2),
\]
The true fixed effects $\beta = (\beta_0, \beta_1, \beta_2, \beta_3, \beta_4)$ are specified by the argument \code{beta}; the true error variance is specified by the argument \code{sigma2}, and \code{covb} specifies the random-effects covariance.  

For the competing risks outcome, the cause-specific hazard for failure type
\(k = 1,2\) was specified as
\[
\lambda_{ik}(t \mid  X_i, b_i)
=
\lambda_{0k}
\exp\left( X_i^\top \gamma_k
+ b_i^\top \alpha_k
\right), \quad k = 1,2,
\]
where \(\lambda_{01}\) and \(\lambda_{02}\) is the constant baseline hazard rate specified by \code{lambda1} and \code{lambda2}, \(\gamma_1\) and \(\gamma_2\) are true survival fixed effects specified by \code{gamma1} and \code{gamma2}, and \(\alpha_1\) and  \(\alpha_2\) denotes the true association parameters specified by \code{alpha1} and \code{alpha2}. The censoring time was independently generated from
\[
C_i \sim \mathrm{Uniform}(C_L, C_U).
\]
The observed survival time was defined as the minimum of the two event times
and the censoring time. The event indicator was coded as 0 for censoring, 1
for failure type 1, and 2 for failure type 2. The argument \code{CR = TRUE} indicates a competing risk to be simulated.

The argument \code{seed} denotes a random seed number for a simulated dataset. Each simulated dataset consists of a longitudinal data frame, denoted by \code{ydata}, and a survival data frame, denoted by \code{cdata}. The longitudinal dataset contains repeated biomarker measurements, visit times, subject identifiers, and baseline covariates, whereas the survival dataset contains one record per subject with the observed event or censoring time, competing-risk event indicator, and baseline covariates.

\subsection{Simulation of three longitudinal biomarkers and competing-risks data}
\label{appen:mvjmcs}
To illustrate the use of \code{mvjmcs()}, we generated a simulated dataset containing three longitudinal biomarkers and competing-risks survival data using \code{simmvJMdata()}. This function assumes a multivariate joint model with correlated subject-specific random effects shared between the longitudinal and event processes. The resulting object \code{mvdat} contains the longitudinal data \code{mvydata} and survival data \code{mvcdata}, which are extracted and used for subsequent model fitting.
\begin{Code}
R> mvdat <- simmvJMdata(
+    seed = 100,
+    N = 5000,
+    beta = list(
+      beta1 = c(5, 1.5, 2),
+      beta2 = c(10, 1, 2, 1),
+      beta3 = c(8, 1.2, 1.5, 0.8, 0.4)
+    ),
+    sigma = rep(1, 3),
+    gamma1 = c(1, 0.5),
+    gamma2 = c(-0.5, 0.5),
+    alpha1 = list(
+      alpha11 = -0.5,
+      alpha12 = c(0.5, 0.7),
+      alpha13 = c(0.3, 0.4, 0.1)
+    ),
+    alpha2 = list(
+      alpha21 = 0.5,
+      alpha22 = c(0.5, 0.8),
+      alpha23 = c(0.3, 0.1, -0.2)
+    ),
+    lambda1 = 0.05,
+    lambda2 = 0.05,
+    covb = diag(c(5, 10, 1, 10, 1, 0.5)),
+    CR = TRUE
+  )
R> mvydata <- mvdat[["mvydata"]]
R> mvcdata <- mvdat[["mvcdata"]]
\end{Code}
For subject \(i=1,\ldots,n\), two baseline covariates
\(X_i=(X_{i1},X_{i2})^\top\) were generated according to
\[
X_{i1} \sim \mathrm{Bernoulli}(0.5), \qquad
X_{i2} \sim \mathrm{Uniform}(-5,5).
\]
The number of longitudinal biomarkers in a simulation is determined by the length of the
list argument \code{beta}. Specifically, if \code{beta} contains \(G\)
components, then \(G\) biomarkers are generated. For biomarker \(g\),
the corresponding fixed-effect parameter vector is specified by
\code{beta[[g]]}, the error variance is specified by
\code{sigma[g]}, and the association parameters for the competing-risks
outcomes are specified by \code{alpha1[[g]]} and
\code{alpha2[[g]]}.
The random-effects structure of each biomarker is determined by the
length of the corresponding association parameter vector. A length of
one corresponds to a random-intercept model, a length of two corresponds
to a random intercept-and-slope model, and a length of three corresponds
to a random intercept, random slope, and random quadratic time-effect
model.

In the simulation considered here, \code{beta} is specified as a list of
three vectors of true fixed-effect parameters, yielding three longitudinal
biomarkers. Biomarker 1 was generated using a random intercept, biomarker
2 was generated using a random intercept and random slope, and biomarker
3 was generated using a random intercept, random slope, and random
quadratic time effect. Consequently, the subject-specific random effects
were generated jointly as
\[
b_i =
(b_{i10}, b_{i20}, b_{i21}, b_{i30}, b_{i31}, b_{i32})^\top
\sim
N(0,\Sigma_b),
\]
where \(b_{i10}\) denotes the random intercept for biomarker 1,
\(b_{i20}\) and \(b_{i21}\) denote the random intercept and random slope
for biomarker 2, and \(b_{i30}\), \(b_{i31}\), and \(b_{i32}\) denote the
random intercept, random slope, and random quadratic time effect for
biomarker 3. The covariance matrix \(\Sigma_b\) is specified by the
argument \code{covb}.

Given the covariates and random effects specified above, the three
longitudinal biomarkers were generated from biomarker-specific linear
mixed-effects submodels:
\begin{eqnarray*}
Y_{i1}(t_{ij})
&=&
\beta_{10}
+\beta_{11}X_{i1}
+\beta_{12}X_{i2}
+b_{i10}
+\epsilon_{i1}(t_{ij}), \cr
Y_{i2}(t_{ij})
&=&
\beta_{20}
+\beta_{21}X_{i1}
+\beta_{22}X_{i2}
+\beta_{23}t_{ij}
+b_{i20}
+b_{i21}t_{ij}
+\epsilon_{i2}(t_{ij}), \cr
Y_{i3}(t_{ij})
&=&
\beta_{30}
+\beta_{31}X_{i1}
+\beta_{32}X_{i2}
+\beta_{33}t_{ij}
+\beta_{34}t_{ij}^2
+b_{i30}
+b_{i31}t_{ij}
+b_{i32}t_{ij}^2
+\epsilon_{i3}(t_{ij}),
\end{eqnarray*}
where
\begin{eqnarray*}
\epsilon_{ig}(t_{ij}) &\sim& N(0,\sigma_g^2), \quad g=1,2,3.
\end{eqnarray*}
The fixed-effect parameter vectors for the three biomarkers are specified
by the list argument \code{beta}, with one component corresponding to each
biomarker, whereas the error variances \(\sigma_1^2\), \(\sigma_2^2\), and
\(\sigma_3^2\) are specified by the vector argument \code{sigma}.

For the competing risks outcome, the cause-specific hazard for failure type
\(k=1,2\) was specified as
\[
\lambda_{ik}(t \mid X_i,b_i)
=
\lambda_{0k}
\exp\left\{
X_i^\top \gamma_k
+
\alpha_{1k} b_{i10}
+
\alpha_{2k}^\top
\begin{pmatrix}
b_{i20}\\
b_{i21}
\end{pmatrix}
+
\alpha_{3k}^\top
\begin{pmatrix}
b_{i30}\\
b_{i31}\\
b_{i32}
\end{pmatrix}
\right\}.
\]
Here, \(\lambda_{01}\) and \(\lambda_{02}\) denote the constant baseline
hazard rates specified by \code{lambda1} and \code{lambda2}, respectively.
The vectors \(\gamma_1\) and \(\gamma_2\) are the survival fixed-effect
parameters specified by \code{gamma1} and \code{gamma2}. The lists
\code{alpha1} and \code{alpha2} specify the association parameters linking
the random effects from the three longitudinal biomarkers to the hazards of
failure types 1 and 2, respectively. The censoring mechanism, event-time generation procedure, and data structures returned by \code{simmvJMdata()} are identical to those described in Appendix~\ref{appen:jmcs} and are therefore omitted for brevity.

\subsection{Illustration of generating the single-biomarker data with heterogeneous within-subject variability and competing risks data}
\label{appen:JMMLSM}
To illustrate the use of \code{JMMLSM()} in Section \ref{sec:expJMMLSM},
we generated data using \code{simJMWSVdata()}, which extends
\code{simJMdata()} by incorporating heterogeneous within-subject (WS)
variability in the longitudinal process. The common arguments, such as
\code{seed}, \code{N}, \code{increment}, \code{beta}, \code{gamma1},
\code{gamma2}, \code{alpha1}, \code{alpha2}, \code{lambda1},
\code{lambda2}, \code{CL}, \code{CU}, and \code{CR}, have the same
interpretations as in Appendix~\ref{appen:jmcs}. We therefore focus below on
the additional parameters specific to the data-generating mechanism.
In particular, \code{tau} specifies the fixed effects in the variance submodel, \code{vee1} and \code{vee2} quantify the association between subject-specific WS variability and the two cause-specific hazards, and \code{covbw} specifies the covariance matrix of the random effects in both the longitudinal mean and variance components.
\begin{Code}
R> data <- simJMWSVdata(
+    seed = 100,
+    N = 1000,
+    increment = 0.4,
+    beta = c(5, 1.5, 2, 1, 2),
+    tau = c(0.5, 0.5, 0.1, -0.2, 0.3),
+    gamma1 = c(1, 0.5, 0.5),
+    gamma2 = c(-0.5, 0.5, 0.25),
+    alpha1 = c(-0.4, 0.7),
+    alpha2 = c(-0.2, -0.5),
+    vee1 = 0.2,
+    vee2 = -0.3,
+    lambda1 = 0.05,
+    lambda2 = 0.1,
+    CL = 5,
+    CU = 10,
+    covbw = diag(c(10, 1, 0.5)),
+    CR = TRUE
+  )
R> ydatah <- data[["ydatah"]]
R> cdatah <- data[["cdatah"]]
\end{Code}
For subject \(i = 1,\ldots,n\), three baseline covariates
\(X_i=(X_{i1},X_{i2},X_{i3})^\top\) were generated:

\[
X_{i1} \sim \mathrm{Bernoulli}(0.5), \qquad
X_{i2} \sim \mathrm{Uniform}(-1,1), \qquad
X_{i3} \sim N(1,2^2).
\]
The subject-specific random effects were generated as
\[
\theta_i = (b_{i0}, b_{i1}, \omega_i)^\top
\sim N(0,\Sigma_{\theta}),
\]
where \(b_{i0}\) and \(b_{i1}\) denote the random intercept and random slope in the longitudinal mean submodel, respectively, and \(\omega_i\) is a subject-specific random effect governing WS variability. The covariance matrix \(\Sigma_{\theta}\) is specified by the argument \code{covbw}.
Given the covariates and random effects, the longitudinal biomarker was generated according to
\[
Y_i(t_{ij})
=
\beta_0
+\beta_1 X_{i1}
+\beta_2 X_{i2}
+\beta_3 X_{i3}
+\beta_4 t_{ij}
+b_{i0}
+b_{i1} t_{ij}
+\varepsilon_i(t_{ij}), \quad
\varepsilon_i(t_{ij})
\sim
N\!\left(0,\sigma_i^2(t_{ij})\right),
\]
\[
\log\{\sigma_i^2(t_{ij})\}
=
\tau_0
+\tau_1 X_{i1}
+\tau_2 X_{i2}
+\tau_3 X_{i3}
+\tau_4 t_{ij}
+\omega_i,
\]
where
\(\tau=(\tau_0,\tau_1,\tau_2,\tau_3,\tau_4)\) is specified by the argument \code{tau}. 

For the competing risks outcome, the cause-specific hazard for failure type
\(k=1,2\) was specified as
\[
\lambda_{ik}(t \mid X_i,b_i,w_i)
= \lambda_{0k} \exp\!\left(X_i^\top\gamma_k + \alpha_k^\top b_i + \nu_k \omega_i\right), \quad k=1,2,
\]
where the parameters \(\nu_1\) and \(\nu_2\), specified by \code{vee1} and \code{vee2}, quantify the association between WS variability and the hazards of failure types 1 and 2, respectively. The censoring mechanism, event-time generation procedure, and data structures returned by \code{simJMWSVdata()} are identical to those described in Appendix~\ref{appen:jmcs} and are therefore omitted for brevity.

\section{Additional details for landmark multivariate joint modeling}
\label{Appen:supp_landmark_joint}
This section provides additional methodological and computational details for the landmark multivariate joint modeling approach implemented in \code{mvjmcs()}, as outlined in Section~\ref{sec:5}. The method combines landmark prediction with multivariate joint modeling among subjects who remain event-free at a pre-specified landmark time $s$. In contrast to conventional landmark models based directly on observed biomarker values, the proposed approach accounts for measurement error and uses the estimated subject-specific biomarker trajectories to characterize their latent associations with event risks.

\subsection{Assumptions}
The landmark multivariate joint model relies on the following assumptions:
\begin{enumerate}
\item The longitudinal mixed-effects submodels adequately describe the biomarker trajectories through the landmark time \(s\).
\item Conditional on the random effects and observed covariates, the longitudinal measurement errors are independent of the post-landmark event process.
\item Censoring after the landmark time \(s\) is noninformative conditional on the included covariates and random effects.
\item The random-effects distribution provides an adequate approximation within the landmark risk set.
\item The landmark time and prediction horizon are selected independently of unobserved future outcomes.
\end{enumerate}

\subsection{Landmark population and observed data}
The risk set for the landmark multivariate joint model is defined as
\[
\mathcal{R}(s)
=
\left\{
i:T_i>s
\right\},
\]
which only includes subjects who remain event-free at the landmark time \(s\). No additional eligibility criterion is imposed on the timing or number of longitudinal measurements, and all available measurements $Y_{i}(t)$ from subjects in $\mathcal{R}(s)$ are used for model estimation. The definitions of \((T_i,D_i)\) are the same as those given in Section~\ref{Sec:2.1}. Let \(\Psi_s\) denote the landmark-specific parameter vector estimated
using data from subjects in \(\mathcal{R}(s)\). Because the model is fitted separately at each landmark time, the parameter estimates
and baseline hazards may vary with \(s\). For notational simplicity, the
subscript \(s\) is suppressed from the individual components of
\(\Psi_s\), including \(\beta_g\), \(\sigma_g^2\), \(\Sigma\), \(\Lambda_0(.)\),
\(\gamma_k\), and \(\alpha_{gk},\; g=1,\ldots,G,\) throughout the following development.

\subsection{Model formulation}
For each biomarker \(g=1,\ldots,G\), the longitudinal process is characterized by the linear mixed-effects submodel
\begin{eqnarray*}
Y_{ig}(t)
&=& m_{ig}(t)+\sigma_g\epsilon_{ig}(t), \quad \epsilon_{ig}(t)\stackrel{\mathrm{i.i.d.}}{\sim}\mathcal{N}(0,1),
\\
&=&
X_{ig}^{\top}(t)\beta_g+
Z_{ig}^{\top}(t)b_{ig} + \sigma_g\epsilon_{ig}(t), 
\end{eqnarray*}
where \(m_{ig}(t)\) is the true mean biomarker trajectory of $Y_{ig}(t), \; \forall i \in \mathcal{R}(s)$. The joint vector of random effects from all biomarkers are given by
\(b_i=
\left(
b_{i1}^{\top},\ldots,b_{iG}^{\top}
\right)^{\top} \sim\mathcal{N}(0,\Sigma)\). 

The cause-specific hazard after the landmark time $s$ is modeled as
\begin{equation}
\lambda_{ik}
\left\{
t\mid W_i(s),b_i
\right\}
=
\lambda_{0k}(t \mid s)
\exp\left\{
W_i(s)^{\top}\gamma_k+
\sum_{g=1}^{G}\alpha_{gk}A_{ig}(s)
\right\},
\; t \in (s, \infty),
\label{eq:supp_landmark_hazard}
\end{equation}
where \(\lambda_{0k}(t \mid s)\) denotes the conditional cause-specific baseline hazard, \(W_i(s)\) is a vector of baseline or landmark covariates, \(\gamma_k\) is the corresponding regression coefficient vector, and $A_{ig}(s)$ is a pre-specified association
function linking the longitudinal process for biomarker \(g\) to the
cause-specific event process. For example, $A_{ig}(s)$ may be defined as the current value of biomarker \(g\) evaluated at the landmark time $s$:
\begin{equation}
\label{eq:supp_association1}
A_{ig}(s)=m_{ig}(s),
\end{equation}
or as the current value of latent process:
\begin{equation}
A_{ig}(s)
= Z_{ig}^{\top}(s)b_{ig},
\label{eq:supp_association2}
\end{equation}
where \(\alpha_{gk}\) represents the association parameter of $A_{ig}(s)$. Both association structures are currently implemented in \pkg{FastJM}. Notably, equation~\eqref{eq:supp_landmark_hazard} is evaluated at the landmark time \(s\) and remains fixed during the prediction interval \(t>s\), which differs from a conventional joint model where the cause-specific hazard $\lambda_{ik}(.)$ at time \(t\) depends on \(A_{ig}(t)\). This landmark formulation~\eqref{eq:supp_landmark_hazard} reduces computational cost while retaining adjustment for measurement error and subject-specific biomarker trajectories.

\subsection{Likelihood and model estimation}
\subsubsection{Likelihood}
Omitting the covariates for the sake of brevity, the full observed-data likelihood conditional on the landmark time $s$ is given by
\begin{eqnarray}
\label{Appen:eq:supp_full_likelihood}
    \prod_{i\in\mathcal{R}(s)} L_i(Y_i, T_i, D_i \mid s; \Psi)= \prod_{i\in\mathcal{R}(s)} \int \left\{\prod_{g=1}^G f(Y_{ig} \mid b_{i}; \Psi)\right\}f(T_i, D_i \mid b_i, s; \Psi)f(b_i; \Psi)\,db_i,
\end{eqnarray}
where 
\begin{eqnarray*}
    f(Y_{ig} \mid b_i; \Psi) &=& \prod_{j=1}^{n_{ig}} \frac{1}{\sqrt{2\pi\sigma_g^2}}\exp \left[-\frac{1}{2\sigma_g^2}\left\{Y_{ig}(t_{ijg})-X_{ig}^{\top}(t_{ijg})\beta_g - Z_{ig}^{\top}(t_{ijg})b_{ig}\right\}^2\right], \\
    f(T_i, D_i \mid b_i, s; \Psi) &=& \prod_{k=1}^K \left[\Delta \Lambda_{0k}(T_i \mid s)\exp\left\{W_i(s)^{\top}\gamma_k+\sum_{g=1}^{G}\alpha_{gk}A_{ig}(s)\right\}\right]^{I(D_i=k)} \cr
    &&\times \exp\left[- \sum_{k=1}^K \Lambda_{0k}(T_i \mid s)\exp\left\{W_i(s)^{\top}\gamma_k+\sum_{g=1}^{G}\alpha_{gk}A_{ig}(s)\right\}\right], \\
    f(b_i; \Psi) &=&  \frac{1}{\sqrt{(2\pi)^{q}\mid\Sigma\mid}}\exp(-\frac{1}{2}b_i^{\top} \Sigma^{-1} b_i),
\end{eqnarray*}
where $\Lambda_{0k}(.)$ is the cumulative baseline hazard function for type $k$ failure and $\Delta\Lambda_{0k}(T_i\mid s)=\Lambda_{0k}(T_i\mid s) - \Lambda_{0k}(T_i-\mid s)$.
Note that the first equality follows from the assumption that $Y_i$ and $(T_i, D_i)$ are independent conditional on the covariates, the random effects, and the landmark time.

\subsubsection{EM algorithm}
As discussed in Section \ref{sec:2.3}, direct maximization of the likelihood~\eqref{Appen:eq:supp_full_likelihood} is computationally challenging, thus a customized expectation-maximization (EM) algorithm can be developed to obtain the (semiparametric) MLE, in which the random effects are treated as missing data. Specifically, the complete-data log-likelihood is maximized by alternating between an E-step,
\begin{equation}
\label{Appen:Estep}
Q(\Psi\mid\Psi^{(m)},s)
=
\sum_{i\in\mathcal{R}(s)} E_{\Psi^{(m)}}\left\{
\log L_i(Y_i,T_i,D_i,b_i\mid s;\Psi)
\right\},
\end{equation}
and the M-step
\begin{equation}
\label{Appen:Mstep}
\Psi^{(m+1)}
=
\arg\max_{\Psi}
Q(\Psi\mid\Psi^{(m)},s).
\end{equation}
These two steps are repeated until the algorithm converges, where
\(\Psi^{(m)}\) denotes the current parameter estimate at iteration \(m\).
The expectation in~\eqref{Appen:Estep} is taken with respect to the
conditional distribution
\[
f(b_i\mid Y_i,T_i,D_i,s;\Psi^{(m)}).
\]

The subject-specific complete-data log-likelihood required in the
E-step is
\begin{equation}
\label{complikelihood}
\log L_i(Y_i,T_i,D_i,b_i\mid s;\Psi)
=
\sum_{g=1}^{G}
\log f(Y_{ig}\mid b_i;\Psi)
+
\log f(T_i,D_i\mid b_i,s;\Psi)
+
\log f(b_i;\Psi).
\end{equation}
When the parameter vector can be partitioned into nonoverlapping blocks
corresponding to the longitudinal, survival, and random-effects
components, the conditional expectation of each term
in~\eqref{complikelihood} can be maximized separately in the M-step.
This maximization requires posterior expectations of certain functions
of the random effects, denoted by \(h(b_i)\). At iteration \(m\), these
quantities are computed in the E-step as
\[
E_{\Psi^{(m)}}\!\left[
h(b_i)\mid Y_i,T_i,D_i,s
\right]
=
\int
h(b_i)
f(b_i\mid Y_i,T_i,D_i,s;\Psi^{(m)})
\,db_i,
\]
and are subsequently used to update the corresponding parameter blocks
in the M-step.

To efficiently compute the expectations involving \(h(b_i)\) in the E-step~\eqref{Appen:Estep}, particularly when \(b_i\) is high-dimensional, \pkg{FastJM} adopts a normal approximation approach developed by \citet{bernhardt2015fast,murray2022fast}, in which the posterior distribution of the random effects given the observed data can be approximated by a multivariate normal distribution, followed by
\begin{eqnarray}
\label{biappro}
    b_i \mid Y_i, T_i, D_i, s;\hat{\Psi} \approx N(\hat{b}_i, \hat{\Sigma}_i),
\end{eqnarray}
where $\hat{b}_i$ is the posterior mode of $f(Y_i, T_i, D_i, b_i \mid s; \hat{\Psi})$ at the current EM iteration, with the estimated covariance
\begin{eqnarray*}
    \hat{\Sigma}_i = \left\{ -\frac{\partial^2 \log L_i(Y_i, T_i, D_i, b_i,s | \hat{\Psi})}{\partial b_i \partial b_i ^{\top}} \Big |_{b_i = \hat{b}_i} \right\}^{-1}.
\end{eqnarray*}
The normal approximation of the conditional distribution of $b_i$ was previously demonstrated in \citet{rizopoulos2012fast} as $n_{ig} \rightarrow \infty$, and \citet{bernhardt2015fast,murray2022fast,murray2023fast} also demonstrated that the parameter estimation appears reasonable even for fewer longitudinal follow-up measures. Following the approximation (\ref{biappro}), any linear combination of $b_i$ is also normal, thus their expectations in the E-step (\ref{Appen:Estep}) can be calculated analytically. 
%in approximation (\ref{biappro}). 
As for the non-linear terms in the E-step (\ref{Appen:Estep}), \citet{li2025efficient} proposed to approximate the expectations of the non-linear terms via moment generating function (MGF). In the following, we describe the normal approximation algorithm used to update the parameters from iteration $(m)$ to iteration $(m +1)$.

\paragraph{Update for \texorpdfstring{$\beta$}{beta}}
We first set the log-likelihood $\log L_i (Y_i, T_i, D_i, b_i \mid s; \Psi)$ with respect to $\beta$,
\begin{eqnarray*}
    \log L_i(\beta) \propto -\frac{1}{2} (Y_i - X_i^{\top}\beta - Z_i^{\top} b_i)^{\top} V_i^{-1} (Y_i - X_i^{\top}\beta - Z_i^{\top} b_i),
\end{eqnarray*}
where $X_i = \bigoplus_{g=1}^G X_{ig}$, $Z_i = \bigoplus_{g=1}^G Z_{ig}$, and $V_i = \bigoplus_{g=1}^G \sigma_g^2 I_{n_{ig}}$, with $\bigoplus$ denoting the direct matrix sum and $I_x$ denoting an $x \times x$ identity matrix.

Now we need to compute the expectation of the above log-likelihood with respect to the posterior distribution of $b_i$.

\[
E[\log L_i(\beta)] = E \left[ -\frac{1}{2} (Y_i - X_i^{\top} \beta - Z_i^{\top} b_i)^{\top} V_i^{-1} (Y_i - X_i^{\top} \beta - Z_i^{\top} b_i) \right]
\]
The random variable is \( b_i \), and the fixed parameter is \( \beta \). Using the approximation (\ref{biappro}), we have the conditional distribution of \( b_i \) given the observed data is:
\[
b_i \mid Y_i, T_i, D_i, s; \Psi^{(m)} \sim N(\hat{b}_i, \hat{\Sigma}_i),
\]
where \( \hat{b}_i \) is the posterior mean, and \( \hat{\Sigma}_i \) is the posterior covariance of \( b_i \). Therefore,

\[Y_i - X_i^{\top} \beta - Z_i^{\top} b_i \mid Y_i, T_i, D_i, \Psi^{(m)} \sim N(Y_i - X_i^{\top} \beta -Z_i^{\top}  \hat{b}_i, Z_i^{\top}\hat{\Sigma}_iZ_i),
\]
Then we have 
\begin{eqnarray*}
E[\log L_i(\beta)] &=& -\frac{1}{2} \left[ 
\text{Tr}\{V_i^{-1}Z_i^{\top}\hat{\Sigma}_iZ_i\}+(Y_i - X_i^{\top} \beta -Z_i^{\top}  \hat{b}_i)^{\top}V_i^{-1}(Y_i - X_i^{\top} \beta -Z_i^{\top}  \hat{b}_i)\right]
\end{eqnarray*}
To find the estimator for $\beta$, we take the derivative of the log-likelihood with respect to $\beta$:
\begin{eqnarray*}
   \frac{\partial E[\log L_i(\beta)]}{\partial \beta} &=&-\frac{1}{2}\frac{\partial}{\partial \beta}(Y_i - X_i^{\top} \beta - Z_i^{\top} \hat{b}_i)^{\top} V_i^{-1} (Y_i - X_i^{\top} \beta - Z_i^{\top} \hat{b}_i)\\
 &=&  -\frac{1}{2}\frac{\partial}{\partial \beta}\left\{(Y_i - Z_i^{\top} \hat{b}_i)^{\top} V_i^{-1} (Y_i - Z_i^{\top} \hat{b}_i)-2 (\beta^{\top}X_i)V_i^{-1}(Y_i - Z_i^{\top} \hat{b}_i) \right. \\
 && \left. + (\beta^{\top} X_i)V_i^{-1} (X_i^{\top} \beta)\right\} \\
&=&\frac{\partial}{\partial \beta} \left[ -\frac{1}{2} (\beta^{\top}X_i) V_i^{-1} ( X_i^{\top} \beta) - \frac{1}{2} \cdot (-2) (\beta^{\top}X_i)V_i^{-1}(Y_i - Z_i^{\top} \hat{b}_i) \right] \\
&=& X_i  V_i^{-1} (Y_i- Z_i^{\top} \hat{b}_i-X_i^{\top} \beta)
\end{eqnarray*}
Setting the derivative to zero:
\begin{eqnarray*}
X_i V_i^{-1} (Y_i -Z_i^{\top} \hat{b}_i) &=& X_i V_i^{-1}X_i^{\top}\beta
\end{eqnarray*}

To derive the estimator for \(\beta\), we sum this equation over all subjects \(i\) from the risk set $\mathcal{R}(s)$. Thus:
\[
\sum_{i\in\mathcal{R}(s)} X_i V_i^{-1} (Y_i - Z_i^{\top} \hat{b}_i) = \sum_{i\in\mathcal{R}(s)} X_i V_i^{-1} X_i^{\top} \beta
\]
Solving for $\beta$ gives:
\[
\hat{\beta} = \left(\sum_{i\in\mathcal{R}(s)} X_i V_i^{-1} X_i^{\top}\right)^{-1} \left\{\sum_{i\in\mathcal{R}(s)} X_i V_i^{-1} \left(Y_i - Z_i^{\top} \hat{b}_i\right)\right\}
\]

\paragraph{\texorpdfstring{Update for $\sigma_g^2$}{Update for sigma\_g²}}
We set the log-likelihood w.r.t. $\sigma_g^2$,
\begin{equation*}
    \log L_i(\sigma_g^2) \propto -\frac{n_{ig}}{2} \log \sigma_g^2 - \sum_{j=1}^{n_{ig}}\frac{1}{2\sigma_g^2} \left\{Y_{ig}(t_{ijg}) - X_{ig}^{\top}(t_{ijg}) \beta_g -Z_{ig}^{\top}(t_{ijg}) b_{ig}\right\}^2.
\end{equation*}
\[
\begin{aligned}
E\!\left[\log L_i(\sigma_g^2)\right]
={}&
-\frac{n_{ig}}{2}\log\sigma_g^2
-\frac{1}{2\sigma_g^2}
\sum_{j=1}^{n_{ig}}
E\!\left[
\left\{
r_{ig}(t_{ijg})
-
Z_{ig}^{\top}(t_{ijg})b_{ig}
\right\}^{2}
\right]
\\
={}&
-\frac{n_{ig}}{2}\log\sigma_g^2
-\frac{1}{2\sigma_g^2}
\sum_{j=1}^{n_{ig}}
\Big[
r_{ig}^{2}(t_{ijg})
-
2r_{ig}(t_{ijg})
Z_{ig}^{\top}(t_{ijg})E(b_{ig})
\\
&+\operatorname{tr}\!\left\{
Z_{ig}(t_{ijg})
Z_{ig}^{\top}(t_{ijg})
E(b_{ig}b_{ig}^{\top})
\right\}
\Big].
\end{aligned}
\]
The first derivative w.r.t. $\sigma_g^2$ is
\begin{eqnarray*}
\frac{\partial E [\log L_i(\sigma_g^2)]}{\partial \sigma_g^2} &=&
 -\frac{n_{ig}}{2\sigma_g^2} + \frac{1}{2\sigma_g^4} \sum_{j=1}^{n_{ig}} \left[r_{ig}^2(t_{ijg}) - 2 r_{ig}(t_{ijg}) Z_{ig}^{\top}(t_{ijg}) E(b_{ig}) + \right.\\
 &&\left. \text{Tr}\left\{Z_{ig}(t_{ijg})Z_{ig}^{\top}(t_{ijg}) E(b_{ig} b_{ig}^{\top})\right\}\right].
\end{eqnarray*}
Equating the above equation to be 0 and solve for $\sigma_g^2$ for all $i=1,\ldots,n$, we obtain the close-form update
\begin{equation*}
    \hat{\sigma}_g^2 =\frac{\sum_{i\in\mathcal{R}(s)} \sum_{j=1}^{n_{ig}} \left[r_{ig}^2(t_{ijg}) -  2 r_{ig}(t_{ijg}) Z_{ig}^{\top}(t_{ijg}) E(b_{ig}) +  \text{Tr}\left\{Z_{ig}(t_{ijg})Z_{ig}^{\top}(t_{ijg}) E(b_{ig} b_{ig}^{\top})\right\}\right]}{\sum_{i\in\mathcal{R}(s)} n_{ig}}. 
\end{equation*}
Using the approximation (\ref{biappro}), we have $E(b_{ig}) = \hat{b}_{ig}$ and $Var(b_{ig}) = \hat{\Sigma}_{ig}$, where $\hat{b}_{ig}$ is the posterior mode w.r.t. $b_{ig}$, i.e., the random effects for the $g^{th}$ longitudinal response, and $\hat{\Sigma}_{ig}$ is the $(g,g)^{th}$ block matrix of $\hat{\Sigma}_i$.  This results in
\begin{eqnarray*}
    E(b_{ig} b_{ig}^{\top}) &=& Var(b_{ig}) + E(b_{ig}) E(b_{ig})^{\top} \\
    &=& \hat{\Sigma}_{ig} + \hat{b}_{ig} \hat{b}_{ig}^{\top}.
\end{eqnarray*}
Therefore, we have 
\begin{equation*}
    \hat{\sigma}_g^2 =\frac{\sum_{i\in\mathcal{R}(s)} \sum_{j=1}^{n_{ig}} \left[r_{ig}^2(t_{ijg}) -  2 r_{ig}(t_{ijg}) Z_{ig}^{\top}(t_{ijg}) \hat{b}_{ig} +  \text{Tr}\left\{Z_{ig}(t_{ijg})Z_{ig}^{\top}(t_{ijg}) (\hat{\Sigma}_{ig} + \hat{b}_{ig} \hat{b}_{ig}^{\top})\right\}\right]}{\sum_{i\in\mathcal{R}(s)} n_{ig}}. 
\end{equation*}

\paragraph{\texorpdfstring{Update for $\Sigma$}{Update for Sigma}}
We have
\[
\begin{aligned}
E\!\left[\log f(b_i\mid\Sigma)\right]
={}&
E\!\left[
-\frac{q}{2}\log(2\pi)
-\frac{1}{2}\log|\Sigma|
-\frac{1}{2}b_i^{\top}\Sigma^{-1}b_i
\right]
\\
={}&
-\frac{q}{2}\log(2\pi)
-\frac{1}{2}\log|\Sigma|
-\frac{1}{2}
\operatorname{tr}\!\left\{
\Sigma^{-1}E(b_i b_i^{\top})
\right\}.
\end{aligned}
\]
The first derivative of $E [\log f(b_i \mid \Sigma)]$ w.r.t. $\Sigma$ is
\begin{equation*}
    \frac{\partial E [\log f(b_i \mid \Sigma)]}{\partial \Sigma} = -\frac{1}{2}\Sigma^{-1} + \frac{1}{2} \Sigma^{-1} E(b_i b_i^{\top}) \Sigma^{-1}.
\end{equation*}
Equating the above equation to be 0, we obtain the close-form update
\begin{eqnarray*}
    \hat{\Sigma} = E(b_i b_i^{\top}).
\end{eqnarray*}
Using the approximation (\ref{biappro}), we have $E(b_i) = \hat{b}_i$ and $Var(b_i) = \hat{\Sigma}_i$. This results in
\begin{eqnarray*}
    E(b_i b_i^{\top}) &=& Var(b_i) + E(b_i) E(b_i)^{\top} \\
    &=& \hat{\Sigma}_i + \hat{b}_i \hat{b}_i^{\top}.
\end{eqnarray*}
Finally, we solve for $\Sigma$ for all $i=1,\ldots,n$
\begin{eqnarray*}
    \hat{\Sigma} = \frac{\sum_{i\in\mathcal{R}(s)} (\hat{\Sigma}_i + \hat{b}_i \hat{b}_i^{\top})}{\sum_{i=1}^n I\left\{i\in\mathcal{R}(s)\right\}}.
\end{eqnarray*}

\paragraph{Update for $\Lambda_{0k}(.)$}
We have
\begin{eqnarray*}
  \log  f(T_i, D_i \mid b_i, s; \Psi) &=& \sum_{k=1}^K \left[I(D_i=k)\left\{\log \Delta \Lambda_{0k}(T_i \mid s) + W_i^{\top} (s) \gamma_k + \sum_{g=1}^G \alpha_{gk}A_{ig}(s)\right\} \right. \cr
  && \left. -  \Lambda_{0k}(T_i \mid s)\exp\left\{W_i^{\top} (s) \gamma_k + \sum_{g=1}^G \alpha_{gk}A_{ig}(s)\right\}\right]  \cr
  &=& \sum_{k=1}^K\left[I(D_i=k)\left\{\log \Delta \Lambda_{0k}(T_i \mid s) + W_i^{\top}(s) \gamma_k + A_{i}^{\top}(s) \alpha_k\right\}\right.  \cr
  && \left.-  \Lambda_{0k}(T_i \mid s)\exp\left\{W_i^{\top}(s) \gamma_k + A_{i}^{\top}(s) \alpha_k\right\}\right],
\end{eqnarray*}
where $A_i(s) = \left(A_{i1}^{\top}(s), \ldots, A_{iG}^{\top}(s)\right)^{\top}$.

We now set the log likelihood w.r.t. $\Delta \Lambda_{0k}(.)$
\begin{eqnarray*}
    \log L_i(\Delta \Lambda_{0k}) &\propto& I(D_i = k) \log \Delta \Lambda_{0k}(T_i \mid s) -  \Lambda_{0k}(T_i \mid s)\exp\left\{W_i^{\top}(s) \gamma_k + A_{i}^{\top}(s) \alpha_k\right\} \\
    E\left[\log L_i(\Delta \Lambda_{0k})\right] &=& I(D_i = k) \log \Delta \Lambda_{0k}(T_i \mid s) -  \Lambda_{0k}(T_i \mid s) E\left[\exp\left\{W_i^{\top}(s) \gamma_k + A_{i}^{\top}(s) \alpha_k\right\}\right].
\end{eqnarray*}
Using the approximation (\ref{biappro}), we have 
\begin{eqnarray}
\label{EWb}
   W_i^{\top}(s) \gamma_k + A_{i}^{\top}(s) \alpha_k \sim N\left(W_i^{\top}(s) \gamma_k + \hat{A}_{i}^{\top}(s) \alpha_k, \alpha_k^{\top} B_i(s) \alpha_k  \right),
\end{eqnarray}
where $B_i(s)$ is the covariance matrix of $\hat{A}_{i}(s)$. For the current value~\eqref{eq:supp_association1} / current value of the latent process parametrization~\eqref{eq:supp_association2} implemented in \pkg{FastJM}, \[B_i(s) = Z_i^{\top}(s) \hat{\Sigma}_i Z_i(s).\]
Based upon equation (\ref{EWb}), calculating $E\left[\exp\left\{A_{i}^{\top}(s) \alpha_k\right\}\right]$ is equivalent to the moment generating function (MGF) of $A_i(s)$, given by $$M_{A_{i}(s)}(t) = \exp\left\{A_{i}^{\top}(s) t + \frac{1}{2}t^{\top} B_i(s) t\right\}$$ evaluated at $t = \alpha_k$. Therefore, we can approximate the expectation above as
\begin{eqnarray*}
    E\left[\log L_i(\Delta \Lambda_{0k})\right] &\approx& \Tilde{E}_i\left[\log L_i(\Delta \Lambda_{0k})\right] \\
    &=& I(D_i = k) \log \Delta \Lambda_{0k}(T_i \mid s) \\
    && -  \Lambda_{0k}(T_i \mid s) \exp\left\{W_i^{\top}(s) \gamma_k + \hat{A}_{i}^{\top}(s) \alpha_k + \frac{1}{2}\alpha_k^{\top} B_i(s) \alpha_k\right\}.
\end{eqnarray*}
Take the first derivative w.r.t. $\Delta \Lambda_{0k}(.)$, we obtain the closed-form update of $ \Lambda_{0k}(.)$
\begin{eqnarray}
\label{Mstep:lambda}
    \hat{\Lambda}_{0k}(t\mid s) = \sum_{l: s< t_{kl} \leq t}  \frac{d_{kl}}{\sum_{r \in R(t_{kl})} \exp\left\{W_r^{\top}(s) \gamma_k + \hat{A}_{r}^{\top}(s) \alpha_k + \frac{1}{2}\alpha_k^{\top} B_r(s) \alpha_k\right\}},
\end{eqnarray}
where $R(t_{kl})$ is the risk set at the uncensored $k$-th failure time $t_{kl}$, and $d_{kl}$ is the number of type $k$ failures at $t_{kl}$, for $k=1, \ldots, K.$

\paragraph{Update for $\gamma_k$ and $\alpha_k$}
Using the survival log-likelihood, we can calculate the expectation on the survival fixed effects $\phi_k = (\gamma_k^{\top}, \alpha_k^{\top})^{\top}$:
\begin{eqnarray*}
    \log L_i(\phi_k) &\propto&
    I(D_i = k) \left\{W_i^{\top}(s) \gamma_k + A_{i}^{\top}(s) \alpha_k\right\} -  \Lambda_{0k}(T_i \mid s) \exp\left\{W_i^{\top}(s) \gamma_k + A_{i}^{\top}(s) \alpha_k\right\}, \cr
    E\left\{\log L_i(\phi_k)\right\} &=&  I(D_i = k) \left[W_i^{\top}(s) \gamma_k + E \left\{A_{i}^{\top}(s) \alpha_k\right\}\right] \\
    && - \Lambda_{0k}(T_i \mid s) \exp\left\{W_i^{\top}(s) \gamma_k\right\} E \left[\exp\left\{A_{i}^{\top}(s) \alpha_k\right\}\right],
\end{eqnarray*}
where $\Lambda_{0k}(.)$ will be substituted with $\hat{\Lambda}_{0k}(.)$ from (\ref{Mstep:lambda}).

We note that there is no closed-form solution for $\phi_k$ and it will be updated via an one-step Newton-Raphson algorithm at iteration $m$ to iteration $m+1$:
\begin{eqnarray*}
    \phi_k^{(m+1)} &=& \phi_k^{(m)} + I_{\phi_k}^{(m)-1}S_{\phi_k}^{(m)},\quad k=1,\ldots,K,
\end{eqnarray*}
where 
\begin{eqnarray}
\label{Mstepscorefunc}
    S_{\phi_k}^{(m)} &=& (S_{\gamma_k}^{(m)T}, S_{\alpha_k}^{(m)T})^{T}, \\
    \label{Mstepinfo}
    I_{\phi_k}^{(m)} &=& 
    \begin{pmatrix}
I_{\gamma_k}^{(m)} & I_{\gamma_k \alpha_k}^{(m)} \\
I_{\gamma_k \alpha_k}^{(m)T} & I_{\alpha_k}^{(m)}
\end{pmatrix}.
\end{eqnarray}
The explicit formulas of all the quantities in (\ref{Mstepscorefunc}) - (\ref{Mstepinfo}) are given by
\begin{eqnarray*}
    S_{\gamma_k}^{(m)} &=& \sum_{i\in\mathcal{R}(s)} \left(I(D_i = k) W_i(s) -  \Lambda_{0k}(T_i \mid s) \exp\left\{W_i^{\top}(s) \gamma_k\right\} E\left[\exp\left\{A_{i}^{\top}(s) \alpha_k\right\}\right]W_i(s)\right), \cr
     S_{\alpha_k}^{(m)} &=& \sum_{i\in\mathcal{R}(s)} \left(I(D_i = k) E \left\{A_{i}(s)\right\} -  \Lambda_{0k}(T_i \mid s) \exp\left\{W_i^{\top}(s) \gamma_k\right\} E\left[A_{i}(s)\exp\left\{A_{i}^{\top}(s) \alpha_k\right\}\right]\right),\cr
     I_{\gamma_k}^{(m)} &=& \sum_{i\in\mathcal{R}(s)} \left(\Lambda_{0k}(T_i \mid s) \exp\left\{W_i^{\top}(s) \gamma_k\right\} E\left[\exp\left\{A_{i}^{\top}(s) \alpha_k\right\}\right]W_i(s)W_i^{\top}(s)\right),\cr
     I_{\alpha_k}^{(m)} &=& \sum_{i\in\mathcal{R}(s)} \left(  \Lambda_{0k}(T_i \mid s) \exp\left\{W_i^{\top}(s) \gamma_k\right\} E\left[A_{i}(s)A_{i}^{\top}(s)\exp\left\{A_{i}^{\top}(s) \alpha_k\right\}\right]\right), \cr
     I_{\gamma_k \alpha_k}^{(m)} &=& \sum_{i\in\mathcal{R}(s)} \left(  \Lambda_{0k}(T_i \mid s) \exp\left\{W_i^{\top}(s) \gamma_k\right\} W_i(s) E\left[A_{i}(s)\exp\left\{A_{i}^{\top}(s) \alpha_k\right\}\right]^{\top}\right).
\end{eqnarray*}
The two expected values $E\left[A_{i}(s)\exp\left\{A_{i}^{\top}(s) \alpha_k\right\}\right]$ and $E\left[A_{i}(s)A_{i}^{\top}(s)\exp\left\{A_{i}^{\top}(s) \alpha_k\right\}\right]$ can be approximated by the first and second derivative of $M_{A_i(s)}(t)$ evaluated at $t = \alpha_k$, respectively:
\begin{eqnarray*}
    E\left[A_{i}(s)\exp\left\{A_{i}^{\top}(s) \alpha_k\right\}\right] &\approx& \exp\left\{\hat{A}_i^{\top}(s) \alpha_k + \frac{1}{2}\alpha_k^{\top} B_i(s) \alpha_k\right\}\left\{B_i(s) \alpha_k + \hat{A}_i(s)\right\},
\end{eqnarray*}
\begin{eqnarray*}
    E\left[A_{i}(s)A_{i}^{\top}(s)\exp\left\{A_{i}^{\top}(s) \alpha_k\right\}\right] &\approx& \exp\left\{\hat{A}_i^{\top}(s) \alpha_k + \frac{1}{2}\alpha_k^{\top} B_i(s) \alpha_k\right\} \cr
    && \times \left[\left\{B_i(s) \alpha_k + \hat{A}_i(s)\right\}\left\{B_i(s) \alpha_k + \hat{A}_i(s)\right\}^{\top} + B_i(s)\right].
\end{eqnarray*}

\subsubsection{Standard error estimation}
Let $\Omega = (\beta, \sigma_1^2, \ldots,\sigma_G^2, \gamma_1, \ldots, \gamma_K, \alpha_{1}, \ldots, \alpha_{K}, \Sigma)$ denote the parametric component of $\Psi$ and $\hat{\Omega}$ its semiparametric MLE. We propose to estimate the variance-covariance matrix of $\hat{\Omega}$ by inverting the empirical Fisher information obtained from the profile likelihood of $\Omega$ \citep{lin2004latent, zeng2005asymptotic, zeng2005simultaneous} as follows:
\begin{equation}
\label{SEestimation2}
\sum_{i\in\mathcal{R}(s)} \left[\nabla_{\Omega} l^{(i)}(\hat{\Omega}; Y, T, D)\right]
\left[\nabla_{\Omega} l^{(i)}(\hat{\Omega}; Y, T, D)\right]^{\top},
\end{equation}
where $\nabla_{\Omega} l^{(i)}(\hat{\Omega}; Y, T, D)$ denotes the observed score vector for the $i$th subject obtained from the profile likelihood $l^{(i)}(\Omega; Y, T, D)$ after profiling out the baseline hazards. Calculating its gradient, however, is difficult because there is no explicit expression for the profile likelihood. Here we approximate $\nabla_{\Omega}l^{(i)}(\hat{\Omega}; Y, T, D)$ by the derivative of the profile expected complete-data log-likelihood, obtained in the last E-step when the EM algorithm has converged. The parametric components of the observed score function $\nabla_{\Omega}l^{(i)}(\hat{\Omega}; Y, T, D)$ in equation (\ref{SEestimation2}) can be calculated via normal approximation. Let $r_{ig}(t_{ijg}) = Y_{ig}(t_{ijg})-X_{ig}^{\top}(t_{ijg})\beta_g$ and $U_{ik}(s)=\hat{A}_i^{\top}(s)\alpha_k + \frac{1}{2}\alpha_k^{\top}B_i(s)\alpha_k$. The observed score vector for each parametric component is given by
{\small
\begin{eqnarray}
\label{Appen:SEbeta}
\nabla_{\beta_g}l^{(i)} (\hat{\Omega};Y, T, D)\!\!\! &=& \!\!\! \frac{1}{\sigma_g^2}\sum_{j=1}^{n_{ig}} E\left\{r_{ig}(t_{ijg}) - Z_{ig}^{\top}(t_{ijg}) b_{ig}\right\} X_{ig}(t_{ijg})\Bigg|_{\Omega = \hat{\Omega}}, \cr
&\approx& \frac{1}{\sigma_g^2} \sum_{j=1}^{n_{ig}} r_{ig}(t_{ijg})X_{ig}(t_{ijg}) - X_{ig}(t_{ijg}) Z_{ig}^{\top}(t_{ijg}) \hat{b}_{ig} \Bigg|_{\Omega = \hat{\Omega}} \\
\label{Appen:SEsigma2}
\nabla_{\sigma_g^2}l^{(i)} (\hat{\Omega};Y, T, D)\!\!\! &=&\!\!\! \left[ \frac{1}{2\sigma_g^4} \sum_{j=1}^{n_{ig}} E\left\{r_{ig}(t_{ijg}) - Z_{ig}^{\top}(t_{ijg}) b_{ig}\right\}^2 \! - \! \frac{n_{ig}}{2\sigma_g^2} \right]\Bigg|_{\Omega = \hat{\Omega}}, \cr
&\approx& \sum_{j=1}^{n_{ig}}\frac{\left[r_{ig}^2(t_{ijg}) -  2 r_{ig}(t_{ijg}) Z_{ig}^{\top}(t_{ijg}) \hat{b}_{ig} +  \text{Tr}\left\{Z_{ig}(t_{ijg})Z_{ig}^{\top}(t_{ijg}) (\hat{\Sigma}_{ig} + \hat{b}_{ig} \hat{b}_{ig}^{\top})\right\}\right]}{2\sigma_g^4} - \frac{n_{ig}}{2\sigma_g^2}, \\
\label{Appen:SESigma}
\nabla_{\Sigma}l^{(i)} (\hat{\Omega};Y, T, D) \!\!\!  &=& \!\!\! \frac{1}{2}\left[2\Sigma^{-1}E(b_i b_i^{\top}) \Sigma^{-1}\!\! -\!\!  \left\{\Sigma^{-1}E(b_i b_i^{\top}) \Sigma^{-1} \circ I\right\} \!\! -\!\! 2\Sigma^{-1}\! + \!\Sigma^{-1} \circ I\right]\Bigg|_{\Omega = \hat{\Omega}}, \cr
&\approx& \frac{1}{2}\left[2\Sigma^{-1} (\hat{\Sigma}_i + \hat{b}_i \hat{b}_i^{\top}) \Sigma^{-1}\!\! -\!\!  \left\{\Sigma^{-1} (\hat{\Sigma}_i + \hat{b}_i \hat{b}_i^{\top}) \Sigma^{-1} \circ I\right\} \!\! -\!\! 2\Sigma^{-1}\! + \!\Sigma^{-1} \circ I\right], \\
\label{Appen:SEgamma}
\nabla_{\gamma_k}l^{(i)}(\hat{\Omega};Y, T, D)\!\!\! &=& \!\!\! I(D_i = k) \left[ W_i(s) - \frac{\sum_{r \in R(T_i)} \exp\left\{W_r^{\top}(s) \gamma_k\right\}  E\left[\exp\left\{A_r^{\top}(s) \alpha_k\right\}\right]W_r(s)}{\sum_{r \in R(T_i)} \exp\left\{W_r^{\top}(s) \gamma_k\right\}  E\left[\exp\left\{A_r^{\top}(s) \alpha_k\right\}\right]} \right] \cr
&& + \left\{ \sum_{j: t_{kj} \leq T_i} \frac{d_{kj}\sum_{r \in R(t_{kj})} \exp\left\{W_r^{\top}(s) \gamma_k\right\}  E\left[\exp\left\{A_r^{\top}(s) \alpha_k\right\}\right]W_r(s)}{\left(\sum_{r \in R(t_{kj})} \exp\left\{W_r^{\top}(s) \gamma_k\right\}  E\left[\exp\left\{A_r^{\top}(s) \alpha_k\right\}\right]\right)^2} \right. \cr
&& \left. -\sum_{j: t_{kj} \leq T_i}\frac{d_{kj}}{\sum_{r \in R(t_{kj})} \exp\left\{W_r^{\top}(s) \gamma_k\right\}  E\left[\exp\left\{A_r^{\top}(s) \alpha_k\right\}\right]}W_i(s) \right\} \cr
&& \times \exp\left\{W_i^{\top}(s) \gamma_k\right\}   E\left[\exp\left\{A_i^{\top}(s) \alpha_k\right\}\right]\Bigg|_{\Omega = \hat{\Omega}}, \cr
 &\approx& \!\!\! I(D_i = k) \left[ W_i(s) - \frac{\sum_{r \in R(T_i)} \exp\left\{W_r^{\top}(s) \gamma_k + U_{rk}(s)\right\} W_r(s)}{\sum_{r \in R(T_i)} \exp\left\{W_r^{\top}(s) \gamma_k + U_{rk}(s)\right\}} \right] \cr
&& + \left( \sum_{j: t_{kj} \leq T_i} \frac{d_{kj}\sum_{r \in R(t_{kj})} \exp\left\{W_r^{\top}(s) \gamma_k + U_{rk}(s)\right\}W_r(s)}{\left[\sum_{r \in R(t_{kj})} \exp\left\{W_r^{\top}(s) \gamma_k + U_{rk}(s)\right\}\right]^2} \right. \cr
&& \left. -\sum_{j: t_{kj} \leq T_i}\frac{d_{kj}}{\sum_{r \in R(t_{kj})} \exp\left\{W_r^{\top}(s) \gamma_k + U_{rk}(s)\right\}}W_i(s) \right) \cr
&& \times \exp\left\{W_i^{\top}(s) \gamma_k + U_{ik}(s)\right\}\Bigg|_{\Omega = \hat{\Omega}}, \\
\label{Appen:SEnu}
\nabla_{\alpha_k}l^{(i)}(\hat{\Omega};Y, T, D) &=& I(D_i = k)\left[ E\left\{A_i(s)\right\} - \frac{\sum_{r \in R(T_i)} \exp\left\{W_r^{\top}(s) \gamma_k\right\} E\left[\hat{A}_r(s)\exp\left\{A_r^{\top}(s) \alpha_k\right\}\right]}{\sum_{r \in R(T_i)} \exp\left\{W_r^{\top}(s) \gamma_k\right\} E\left[\exp\left\{A_r^{\top}(s) \alpha_k\right\}\right]}\right] \cr
&& + \left\{ \sum_{j: t_{kj} \leq T_i} \frac{d_{kj}\sum_{r \in R(t_{kj})} \exp\left\{W_r^{\top}(s) \gamma_k\right\} E\left[\hat{A}_r(s)\exp\left\{A_r^{\top}(s) \alpha_k\right\}\right]}{\left(\sum_{r \in R(t_{kj})} \exp\left\{W_r^{\top}(s) \gamma_k\right\} E\left[\exp\left\{A_r^{\top}(s) \alpha_k\right\}\right]\right)^2} E\left[\exp\left\{A_i^{\top}(s) \alpha_k\right\}\right] \right. \cr
&& \left. - \sum_{j: t_{kj} \leq T_i}\frac{d_{kj}}{\sum_{r \in R(t_{kj})} \exp\left\{W_r^{\top}(s) \gamma_k\right\} E\left[\exp\left\{A_r^{\top}(s) \alpha_k\right\}\right]} E\left[A_i(s)\exp\left\{A_i^{\top}(s) \alpha_k\right\}\right] \right\} \cr
&& \times \exp\left\{ W_i^{\top}(s) \gamma_k\right\}\Bigg|_{\Omega = \hat{\Omega}}, \cr
&\approx& I(D_i = k)\left[\hat{A}_i(s) - \frac{\sum_{r \in R(T_i)} \exp\left\{W_r^{\top}(s) \gamma_k + U_{rk}(s)\right\}\left\{B_r(s) \alpha_k +\hat{A}_r(s)\right\}}{\sum_{r \in R(T_i)} \exp\left\{W_r^{\top}(s) \gamma_k + U_{rk}(s)\right\}}\right] \cr
&& + \left( \sum_{j: t_{kj} \leq T_i} \frac{d_{kj}\sum_{r \in R(t_{kj})} \exp\left\{W_r^{\top}(s) \gamma_k + U_{rk}(s)\right\}\left\{B_r(s) \alpha_k +\hat{A}_r(s)\right\}}{\left[\sum_{r \in R(t_{kj})} \exp\left\{W_r^{\top}(s) \gamma_k + U_{rk}(s)\right\}\right]^2} \right. \cr
&& \times \left. \exp\left\{U_{rk}(s)\right\} \right. \cr
&& \left. - \sum_{j: t_{kj} \leq T_i}\frac{d_{kj}}{\sum_{r \in R(t_{kj})} \exp\left\{W_r^{\top}(s) \gamma_k + U_{rk}(s)\right\}} \right.\cr
&& \times \left. \exp\left\{U_{ik}(s)\right\}\left\{B_i(s) \alpha_k + \hat{A}_i(s)\right\} \right)  \exp\left\{ W_i^{\top}(s) \gamma_k\right\}\Bigg|_{\Omega = \hat{\Omega}}. 
\end{eqnarray}
}

\subsection{Dynamic prediction}
For a new subject \(i^*\) with longitudinal history
\[
Y_{i^*}^{(s)}=\{Y_{i^*}(t):t\le s\},
\]
who remains event-free at landmark time \(s\), the cumulative incidence probability of a type-\(k\) failure by a prediction horizon \(u>s\) is
\begin{equation*}
\pi_{i^*k}(u,s)
=
\int
F_{i^*k}(u,s\mid b_{i^*};\Psi)
p\!\left(
b_{i^*}\mid Y_{i^*}^{(s)},T_{i^*}>s;\Psi
\right)
\,db_{i^*}.
\end{equation*}
Here,
\begin{equation*}
F_{i^*k}(u,s\mid b_{i^*};\Psi)=\int_s^uS_{i^*}(v\mid s,b_{i^*};\Psi)\lambda_{i^*k}\!\left(v\mid s,W_{i^*}(s),b_{i^*};\Psi\right)\,dv
\end{equation*}
is the conditional cause-\(k\) cumulative incidence over the prediction window \((s,u]\), where
\[
S_{i^*}(v\mid s,b_{i^*};\Psi)
=
\exp\!\left\{
-\sum_{k=1}^{K}
\int_s^v
\lambda_{i^*k}\!\left(
r\mid s,W_{i^*}(s),b_{i^*};\Psi
\right)
\,dr
\right\}.
\]

The posterior distribution of \(b_{i^*}\), given the subject’s longitudinal history and event-free status at \(s\), is
\begin{equation*}
p\!\left(
b_{i^*}\mid Y_{i^*}^{(s)},T_{i^*}>s;\Psi
\right)
\propto
f\!\left(Y_{i^*}^{(s)}\mid b_{i^*};\Psi\right)f(b_{i^*};\Psi).
\end{equation*}
A plug-in estimate of \(\pi_{i^*k}(u,s)\) may be obtained as
\[
\widehat{\pi}_{i^*k}(u,s)
=
F_{i^*k}(u,s\mid\widehat b_{i^*};\widehat\Psi),
\]
where \(\widehat b_{i^*}\) is the posterior mean of \(b_{i^*}\), and \(\widehat\Psi\) is a vector of the estimated model parameters. Predictions may be updated at a later landmark time \(s'>s\) by reconstructing the risk set and refitting the joint model using the longitudinal information available through \(s'\).

\end{document}